%% file: Havemann2020topics-v21.tex
\documentclass[a4paper,10pt,twocolumn,twoside]{article}
\usepackage{amsmath, amsthm, amssymb}
\usepackage[english]{babel}
\usepackage{chicago}
\usepackage[pdftex]{graphicx}
\usepackage{hyperref}
\usepackage{algorithmic}
\usepackage{algorithm}
\usepackage[utf8]{inputenc}
\usepackage{geometry}
\usepackage{longtable}
\usepackage{booktabs} 
\usepackage{soul}
\usepackage{color} 
\usepackage{array} 
\usepackage{tabularx} 
\hypersetup{
pdfauthor=Frank Havemann,
pdftitle=Topics as clusters of citation links to highly cited sources: the case of research on international relations,
colorlinks=true,
linkcolor=black,
anchorcolor=black,
citecolor=blue,
urlcolor=black,
menucolor=black
}

\title{Topics as Clusters of Citation Links to Highly Cited Sources\\{\large The Case of Research on International Relations}}

\author{Frank Havemann\footnote{Institut f\"{u}r Bibliotheks- und Informationswissen\-schaft, Humboldt-Universit\"at zu Berlin, D 10099 Berlin, Doro\-theen\-str.\;26 (Germany), frank.havemann(at)ibi.hu-berlin.de, ORCID 0000-0002-0485-2580} 
}
\date{}
\begin{document}

\maketitle

\begin{abstract}

Following Henry Small in his approach to co-citation analysis,  highly cited sources are seen as \textit{concept symbols} of research fronts. But instead of co-cited sources I cluster citation links, which are the thematically least heterogenous elements in bibliometric studies.  To  obtain  clusters representing  topics characterised by concepts I restrict link clustering  to citation links to highly cited sources.  
Clusters of citation links between papers in a political-science subfield (International Relations) and 300 of their sources most cited  in the period 2006--2015 are constructed by a local memetic algorithm. It finds local minima in a cost landscape corresponding to  clusters, which can overlap each other pervasively. The  clusters obtained are well separated from the rest of the network but can have suboptimal cohesion. Cohesive cores of topics are found by applying an algorithm that constructs core-periphery structures in link sets. In this methodological paper I only  discuss some first clustering results for the second half of the 10-years period.

\textbf{Keywords:} topics, citation networks, concept symbols, overlapping communities, link clustering, memetic algorithm, core-periphery structures
\end{abstract}

\section{Introduction}
\label{intro}
If a topic is defined as a focus on scientific knowledge shared by a number of researchers then topics should manifest themselves in clusters of co-cited sources, because cited sources represent theoretical, methodological or empirical knowledge used or at least discussed by citing authors. 

Topics can overlap in papers and even more in books 
if they deal with more than one topic. 
Another kind  of overlap can occur on the level of content of topics: shared knowledge itself can be in the foci of researchers working on different topics.   
We therefore need a clustering algorithm that delivers overlapping clusters.

When topics are represented as disjoint clusters of co-cited sources then they
overlap in papers that cite sources in different clusters.  But a cited source can also correspond to more than one topic. We therefore have to allow for overlapping clusters of cited sources, which to the best of my knowledge have not been produced in any co-citation analysis so far. 

Co-citation analysis was independently proposed by Irina \citeN{marshakova1973ssm} and by Henry \citeN{Small1973cocitation}.  \citeN{small1978cited} also introduced the notion of \emph{concept symbols}  represented by highly cited sources, for which co-citation clusters are constructed. 
By adding the papers that cite concept symbols in a co-citation cluster we augment the picture of the corresponding \emph{research front} \cite{garfield_history_1985}.
Co-citation analysis is the usual approach to clustering concept symbols in citation networks but not the only possible one.
 I propose, instead, to cluster citation links from papers to  concept symbols. Link clustering in the bipartite network of citing papers and cited sources avoids the projection onto the co-citation graph of sources and any need for normalising  and  thresholding co-citation strength.  
From clusters of citation links between papers and sources, overlapping clusters of citing research-front papers and of cited concept symbols can be deduced. Thus, we obtain overlapping clusters of highly cited sources that are connected through papers that co-cite them.%
\footnote{Note, that here link clustering is not applied to co-citation links between concept symbols but to citation links in the bipartite network of citing papers and cited sources.\label{fn.bip}}

Among several clustering methods that allow for overlapping clusters,  link clustering has an important advantage when applied to citation networks: citation links are the thematically least heterogenous elements in bibliometric studies. In nearly all cases, a paper cites a source due to only one knowledge claim. Even when a paper refers to 
two or more knowledge claims in a cited source  they often belong to one topic, especially if we search for larger and more general topics as is done here by restricting  link clustering to citation links between papers and highly cited sources.

The topic definition and the link clustering approach applied here have recently been discussed by \citeN{havemann_memetic_2017}. In that paper,  a new evaluation function for link clusters, $\Psi$, and a  local memetic algorithm for link clustering based on this function, PsiMinL,  were proposed and tested for two  kinds of citation networks: a network of direct citations in a set of astronomy papers published within eight years, and a bipartite network of one volume of these papers and all their cited sources. I here also apply PsiMinL to a bipartite network of papers and sources but  restrict the set of sources to highly cited ones. 

Clustering links in networks instead of nodes had been introduced by \citeN{evans2009line} and by \citeN{ahn2009link}. In both approaches graphs are partitioned into disjoint clusters of links. From them overlapping clusters of nodes are deduced. In contrast to these global methods, PsiMinL evaluates each link cluster in a local manner independently from other clusters. It therefore can produce clusters that overlap each other pervasively, i.e., not only in their boundaries but also in inner links and nodes.  A local evaluation of clusters also matches the local character of topics \shortcite{havemann_memetic_2017}.

Clusters or communities in networks are considered as highly cohesive subgraphs that are well separated from the rest of the network \cite{fortunato2010community}. There are cases where these two features of communities cannot be maximised at the same time. Methods can be classified with regard to producing well separated or well connected communities \cite{Rosvall2019dacd}.   Like several other algorithms, PsiMinL 
delivers   clusters that can have low cohesion, i.e., they can easily be split into two or more sub-clusters. This bias of the algorithm is one of the evaluation function $\Psi(L)$ for a cluster given as a link set $L$: it measures separation and is much less sensitive for changes in cohesion  \cite{havemann_communities_2019}. 

The evaluation function $\Psi(L)$ allows for lowly cohesive clusters but that does not hinder its use  for an evaluation of topic clusters.  
Not all knowledge in a shared focus has to be cited in all papers that contribute to the corresponding topic. Only those sources have to be cited that are used for the production of new knowledge. 
Although authors often cite other sources, too, we cannot expect that all sources in a cluster are cited in all papers contributing to the topic. 

Clusters of highly cited sources that represent topics have to be well separated but can have low internal cohesion. 

A second argument for favouring well separated clusters is the hierarchical structure of sets of topics. A topic can have sub-topics, i.e., the splitting of its cluster should not be too difficult. Two topics can also overlap in one sub-topic. Then we have no strict hierarchy but a poly-hierarchy \shortcite{havemann_memetic_2017}. 

Nonetheless, we are interested in cohesive cores of topics corresponding to dense subgraphs of citation networks that are not necessarily well separated from the rest of the network. To extract such dense cores from a well separated link cluster an algorithm was proposed recently by \shortciteN{havemann_communities_2019}. The CPLC-algorithm finds
\underline{c}ore-\underline{p}eriphery structures of \underline{l}ink \underline{c}lusters.      

The analysis reported 
here was made within the Global Pathways project.\footnote{ \url{http://t1p.de/globalpathways}} The aim of this project is to identify topic based, language based and regional or national substructures in research on international relations (IR).

I must leave all conclusions regarding the content structure of IR research to  a forthcoming  paper enriched with the project team's IR competence 
\cite{Risse2020IR}.  I here present results of a test of the proposed approach. 
The focus of the paper is on methodological challenges.

%%%%%%%%%%%%%%%%%%%%%%%%%
\section{Data}
%%%%%%%%%%%%%%%%%%%%%%%%%

For the analysis of IR literature within the Global Pathways project, we wanted to obtain a set of papers in Web of Science (WoS) that prioritises recall over precision. The time span for all downloads was 2006--2015. We started from 115 journals indexed in the WoS category \emph{International Relations} and added  four journals from \emph{Political Science}. 
In the following, these journals are referred to as IR journals. 
We also searched for book chapters
in the Book Citation Index of WoS that are categorised as \emph{International Relations}. 

All documents of those types that are usually published to communicate new research results, namely articles, letters, and proceedings papers (\emph{original papers}), were downloaded, in addition also  reviews and book reviews. WoS also offers access to SciELO (\textit{Scientific Electronic Library Online}, a database mainly covering publications from Latin American countries). From SciELO, records categorised as \emph{International Relations} were downloaded, too. 
The list of journals (Table \ref{tab.journals} on p.\;\pageref{tab.journals})  and further details of data  can be found in Appendix (p.\;\pageref{app}).

After identifying references automatically, as described in Appendix,  the 300 most highly cited sources were selected. I searched manually for further references in the data set that could be identified with them. Here references to different pages  and  editions of books were  identified. The list of top-300 sources can be found in Appendix (Table \ref{e.tab.1} on p.\;\pageref{b.tab.1}). 203 of them have been classified as dealing with IR themes (Table \ref{tab.top.300.distr} on p.\;\pageref{tab.top.300.distr} in Appendix).  
Experiments with clustering smaller numbers of concept symbols revealed that approaching 300 highly cited sources, only peripheral topics were added and  the central topic clusters had become stable.  

%%%%%%%%%%%%%%%%%%%%%%%%%
% methods
%%%%%%%%%%%%%%%%%%%%%%%%%

\section{Methods}
In the following I will discuss some essential elements of the two algorithms applied. Reading these sections is useful for understanding the design of the experiments and their results. Readers who are not interested in  
methodological details can skip the next sections and proceed with the results in section \ref{sec.res}. Further details can be found in the two papers mentioned  \shortcite{havemann_memetic_2017,havemann_communities_2019}.

\subsection{Link clustering: \\PsiMinL algorithm} 

In one sentence, PsiMinL is an evolutionary algorithm that searches in a cost landscape for local minima that correspond to well separated link clusters. Because genetic operators (mutation, crossover, and selection) are combined with deterministic local searches in the cost landscape, PsiMinL can be called a memetic algorithm \cite{neri2012handbook}. A PsiMinL glossary is in the Appendix (p.\;\pageref{sec.gloss}).

Each possible link set $L$ corresponds to a place in the cost landscape, the height of place $L$ is given by the cost function \textit{normalised node-cut} $\Psi(L)$ of link set $L$. A lower value of  $\Psi(L)$ signals a better separated link set $L$.  Normalised node-cut can be defined as
\begin{equation}
\label{eq.Psi}
\Psi(L)= \frac{\sigma(L)}{k_\mathrm{in}(L)} + \frac{\sigma(L)}{k_\mathrm{in}(E-L)}, 
\end{equation} 
with
\begin{equation}
\sigma(L)    = \sum_{i = 1}^n\frac{k_i^\mathrm{in}(L)  (k_i - k_i^\mathrm{in}(L))}{k_i},
\end{equation} 
where $k_i$ is the degree of node $i$, $k_i^\mathrm{in}(L)$ its internal degree with respect to link set $L$, and $k_\mathrm{in}(L)=\sum_{i = 1}^n k_i^\mathrm{in}(L)=2|L|$.
Index $i$ runs through all $n$ nodes but  $k_i^\mathrm{in}(L)=0$ for all nodes that are not attached to a link in $L$. Set $E$ includes all  $m$ edges.  
Note, that $\sigma(L)=\sigma(E-L)$ because $k_i^\mathrm{in}(E-L)=k_i - k_i^\mathrm{in}(L)$, and that $k_\mathrm{in}(E-L)=2m-k_\mathrm{in}(L)$. Thus, $\Psi(L) = \Psi(E-L)$, the cost function of set $L$ equals that of its complement $E-L$.

Deriving their link-clustering approach, \citeN{evans2009line} introduced a random link-node-link walker. The first summand on the right-hand side of Equation \ref{eq.Psi} is the probability of such a walker  sitting on a link in $L$ to escape from $L$ and the second summand is the escape   probability for the complement of $L$ \shortcite{havemann_communities_2019}. 
Further motivations for using the $\Psi$-function were given by \shortciteN{havemann_memetic_2017}.  

A connected link set $L$ that corresponds to a local minimum in the cost landscape is called a link cluster or a link community. The cost landscape is very rough, i.e., there are many local minima that differ only in a few links. We are interested in well separated link sets that differ from any better separated set in more than only some links.   Therefore we need a resolution parameter $r$. It is used to decide whether we can consider a link set $L$ as a \textit{valid} community. If there is a link set $L_0$ with $\Psi(L_0)<\Psi(L)$ and the two link sets differ in less than $r|L|$ links then $L_0$ makes $L$  \textit{invalid}. In other words, we search for local minima with no lower place in the landscape within a radius $r|L|$.  

A local search in PsiMinL is done by greedily including  neighbouring links to a connected link set $L$ or by excluding links  from $L$ that are attached to boundary nodes. Here I have implemented a procedure that tries to lower cost in an alternating sequence of link exclusion and inclusion until no further improvement is possible.    

A simple local search---done by going downhill in the cost landscape---is soon trapped in the next local minimum. We allow the greedy algorithm  in local searches to proceed even when the costs are rising. It stops and goes back to the place $L_\mathrm{min}$ of the last cost minimum in the search if it does not find a place with lower cost after $r|L_\mathrm{min}|$ steps, i.e., if it does not find a link set which makes $L_\mathrm{min}$ invalid. In other words, the local search can tunnel through barriers in the cost landscape if the end of the tunnel is not too far. Then the link set at the end of the tunnel invalidates the cluster at the tunnel entry.\footnote{In Figure \ref{fig:local.search} (in Appendix on p.\;\pageref{fig:local.search}) a cost-size diagram of a local search visualises how the sequence of greedy exclusion and inclusion of links proceeds and how the search path  tunnels through barriers in the cost landscape.} 

In memetic algorithms deterministic local searches are combined with evolutionary genetic operators, i.e., with mutation, crossover, and selection. We need randomness because even  tunneling does not avoid trapping of  local searches in local minima corresponding to invalid communities.  A population is initialised from a seed subgraph by a local search followed by mutations and again local searches until the desired number of different individuals is reached.    
Mutation and crossover are used to explore the cost landscape around a  preliminarily valid cluster at a local minimum that corresponds to the current best individual of a population. 

If two clusters have well separating boundaries their intersection and their union  could also have such a boundary. Therefore, offspring is made from  intersection and union of parents. As one parent the current best individual is chosen, the other one is selected among those individuals that have large genetic distance (measured as set difference) from the best individual. After mutations and crossovers (both followed by local searches) the best individuals are selected for the next generation. 

The memetic algorithm PsiMinL was implemented as an R-package\footnote{The yet unpublished R-package PsiMinL (pro\-gram\-med by Andreas Prescher) and a detailed description of it and its installation will be delivered by request.} with parallel procedures for all members of a population that undergoes an evolution. Because each cluster is evaluated independently from all other ones, several evolutions starting from different seed subgraphs can run parallel, too. As seed subgraphs one can use clusters obtained from any fast clustering algorithm. 
The  set of all valid clusters is totally independent from the set of seeds used to find them but there is no guarantee for finding all valid clusters with a given seed set. 

Different runs of PsiMinL starting from the same seed can end in different local minima of the cost landscape. 
Tests of PsiMinL on the cost landscape of a large citation network of eight years of astronomy papers \shortcite{havemann_memetic_2017} show two typical cases of path bifurcation. The algorithm can run into different hollows, or it ends at different places in the same hollow. In the second case,  distances between different minima were found to be small, often much smaller than the resolution radius $r|L|$. That means, we can assume that further runs of PsiMinL improve and change a result only slightly.

To keep the overview over the many experiments necessary for finding as much as possible valid clusters in a network, it is convenient to ensure that in a local search starting from a mutant or from an offspring of the current best cluster $L_0$ and ending in a better one, $L_0$ is invalidated. 
Consequently, if the first place on the path downhill with a cost $\Psi<\Psi(L_0)$ is not within a radius $r|L_0|$, then the local search is stopped and the individual 
link set
is excluded from further evolution.  

PsiMinL has many parameters (population size, mutation variances and rates, number of crossovers, etc.)\,\,but only resolution $r$ influences the results. All other parameters  only have
influence on the time needed to obtain them.\footnote{Table \ref{tab:parameters} (on p.\;\pageref{tab:parameters} in Appendix) lists parameters, their meanings, and their values chosen in the experiments described below.}

 Recently, \citeN{gabardo2020m} have proposed a new memetic algorithm for global link clustering resulting in overlapping communities of nodes. They evaluate  whole disjoint link partitions with the density metric proposed by \shortciteN{ahn2009link}. 
\citeN{chalupa2018hybrid} have tested different crossover operators combined with deterministic and randomised variants of local search for finding bottle\-necks in networks that correspond to minima of conductance $\Phi$, an evaluation function that favours well separated subgraphs in the world of node clustering as normalised node-cut $\Psi$ does for link clustering.  They found ``sparse imbalanced cuts into a community and the rest of the network, as well as relatively balanced partitions'' (p.\;28 in preprint version). Like that of \citeN{lu_hybrid_2020}  but in contrast to PsiMinL, their algorithm randomly  selects genes of parents for offspring clusters and applies mutation only for population initialisation.  Further papers  related to algorithm PsiMinL are referred to by \shortciteN[p.\;1095]{havemann_memetic_2017}. Evolutionary algorithms used for detecting communities in networks have been reviewed by Clara \citeN{pizzuti_evolutionary_2017}.

Like conductance $\Phi$,  normalised node-cut $\Psi$ neglects the direction of links. Thus, applying it to a bipartite network of papers and their cited sources means that papers and sources are treated symmetrically. 

%%%%%%%%%%%%%%%%%%%%%%%%%%%%%%%%%%
\subsection{Cores and peripheries of link clusters: CPLC algorithm} 
%%%%%%%%%%%%%%%%%%%%%%%%%%%%%%%%%%
CPLC constructs core-periphery structures  (named \textit{towns}, for short) in a given link set  as nested subgraphs with decreasing cohesion. 
Large star subgraphs  have a high local  density of links. 
This density notion is the translation of usual graph density into the world of link clustering \shortcite[p.\;5]{havemann_communities_2019}. For a recent review of algorithms for core-periphery construction see the paper by \citeN{tang_recent_2019}.

In our case the largest stars are highly cited sources with their incoming citation links.  A town is defined as a size ordered cluster of stars where two stars are never indirectly connected via smaller stars only. To illustrate this definition, we can imagine size of stars as height of hills. Then all smaller stars of a town can be reached from the largest one on a path that is never going uphill.  

A star is connected to a town if it shares  a minimum number of outer nodes with the set of town stars of equal or larger size; otherwise it becomes the centre of an independent town. The minimum number of outer nodes is determined by a resolution parameter $q$ with $0 \le q < 1$, which is used as a minimum threshold of relative overlap for a star to be attached to a town. 

Instead of arbitrarily setting parameter $q$, its  whole range is explored  by starting with minimal resolution $q = 0$ and increasing it recursively  to a value at that it is possible to obtain at least one more town in the given link set. To chose a resolution level at which useful core-periphery structures are constructed, different criteria can be applied. One can, e.g.,  consider towns at a level where the two largest stars in the link set are centres of different towns. 

Towns of clusters can also be used to construct appropriate small seed subgraphs for PsiMinL.

%%%%%%%%%%%%%%%%%%%%%%%%%
% experiments
%%%%%%%%%%%%%%%%%%%%%%%%%

\section{Experiments}
\label{sec.exp}
\subsection{Link clustering}
\label{sec:exp.link}

I divided the period 2006--2015 into two 5-years periods for two reasons.
First, because five years are enough to diminish the influence of random fluctuations of citation data. Second, because then a comparison of the two 5-years periods can be made.\footnote{In addition, IR experts can better compare clusters obtained for this period with  the results obtained by \citeN{Kristensen2018IR-end} who analysed author co-citation in IR-papers published in 2011--2015.} 

Any paper that  cites only one of the top-300 sources can be neglected when clusters of them are constructed.  For clustering citation links to these sources, PsiMinL only needs papers that cite at least two of them. 
For 2006--2010 there are  
4,778 such papers and 6,494 papers for the last five years. 
Only papers in IR journals and books were included. 

Seed subgraphs were made from disjoint clusters of cited sources that have been obtained by applying Ward clustering to the co-citation network of top-300 sources. Distances were calculated from the \emph{similarity of views}  \cite{Glaeser2015ISSI}.\footnote{Ward clustering of views was made by Michael Heinz. Its results  can be downloaded as R-object cv.RObj from \url{https://zenodo.org/record/4181930} \cite{havemann_frank_2020_4181930}.} 

%%%%%%%%%%
% figure workflow
%%%%%%%%%%%%%
\begin{figure}[t]
\begin{center}
\includegraphics[height=10cm]{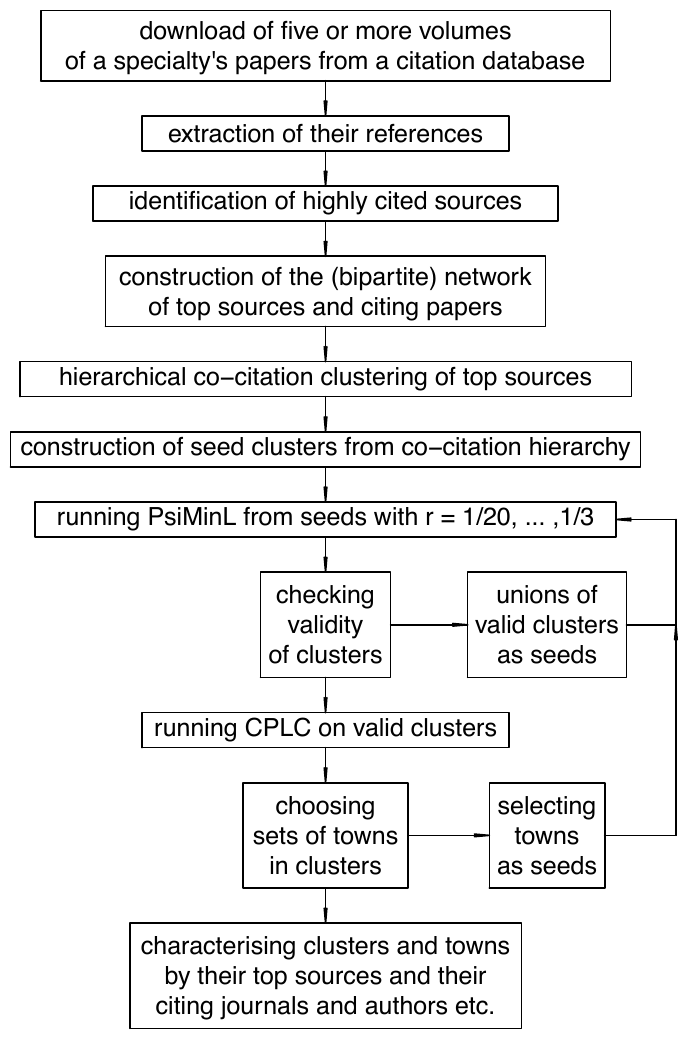}
\caption{Workflow}
\label{fig:workflow}
\end{center}
\end{figure}
%%%%%%%%%%%%%

Usually, an optimal cut through the whole dendrogram of a hierarchical clustering is chosen to get a partition of a network. I have tested this approach to seed construction but starting from 15 middle-sized seeds,  most of  evolutions had a long path to go through the cost landscape: resulting clusters have sizes very different from their seeds (cf.\;Appendix, p.\;\pageref{Ward-15}).  Clusters of one cut through the dendrogram are not well suited as seed subgraphs for an algorithm that results in a poly-hierarchy of clusters. Therefore I have applied an alternative method:  for different numbers of clustered top-300 sources, Ward clusters with longest branches in the dendrogram were selected for constructing seed subgraphs for link clustering.  A Ward cluster has a long branch if it has relatively low variance and if the next larger cluster in the hierarchy has clearly larger variance. Low variance means strong cohesion, large variance of the next super-cluster means weak cohesion or the chance that its sub-clusters are well separated from each other.\footnote{Branch length  measures  cluster quality \cite[p.\;8]{havemann_identifying_2012}.}
The selection of 27 Ward clusters for seed construction is described in Appendix (p.\;\pageref{app:seeds}).\footnote{In addition, the set of seeds has been extended by including further 23 Ward clusters with shorter branches in the dendrogram. Results are in Appendix on p.\;\pageref{23-seeds}.\label{shorter}} 

For any selected Ward cluster of co-cited sources the  set of  citation links to all its sources was used as a seed subgraph for link clustering. PsiMinL first makes a deterministic local search  starting from a seed and then an evolutionary search. 
For a second run of memetic search I made additional seed subgraphs from intersections and unions of valid clusters. 
I also used selected core-periphery structures constructed by applying CPLC on valid clusters as seed subgraphs. 

In all previous experiments we had fixed the resolution parameter on one level: $r=1/3$.
Here I allowed for several levels of resolution. First, resolution parameter $r=1/20$ was chosen, which separates all clusters that differ in at least 1/20 of their links. 
For each seed, 16 independent  evolutions were started with populations of eight individual connected subgraphs given as link sets. 
An evolution was stopped when during 100 generations the best individual could not be improved. In the next phase, the eight best of 16 resulting individuals formed a new population. This was repeated until most of the 16 evolutions gave the same result.\footnote{The parameters used are listed in Table \ref{tab:parameters}  (in Appendix, on p.\;\pageref{tab:parameters}). They had been proven as suitable in a series of previous experiments but until now  no systematic exploration of the parameter space of PsiMinL was made.} 

Then, the whole procedure was repeated  but now with a larger resolution parameter $r$ and using the results of the first run as seeds.  I made such iterations on resolution levels with $r = 1/10, 1/5, 1/4, 1/3$. 
In each step of iteration  a stronger condition for validity was applied than in the step before. All valid link clusters for, e.g., $r=1/4$ are also valid for $r=1/5$ but not the other way round.
\label{sec.res}

The  workflow  of the whole procedure including pre- and post-processing is visualised in Figure \ref{fig:workflow}. The details of the PsiMinL algorithm have been notated as pseudocode by \shortciteN[p.\;1094]{havemann_memetic_2017}.
To give an impression of memetic evolution, the search path starting from a large seed is described and visualised in Appendix (p.\;\pageref{sec:seed.296}).

Figure \ref{fig:cost.size.27} shows  costs $\Psi$ and sizes  of all 27 selected Ward seed-subgraphs, of results of initial local searches and of memetic searches on intermediary resolution levels, and of 11 resulting clusters on final resolution level  ($r=1/3$). 
Each seed is connected by  a line with its intermediary results and its final cluster. The colours of lines  are equal for all evolutions with the same final cluster.  
Cluster $\mathrm{\mathbf{L}}$ (the largest one), e.g., is reached by starting memetic evolution from two large seeds with identifiers 297 and 298 (cf.\;Figure \ref{fig:branch.length} in Appendix, p.\;\pageref{fig:branch.length}). Seed 298 is the largest seed (175 of top-300 sources) and includes seed 297 (103 sources, s.\;Figure \ref{fig:dendro}  in Appendix, p.\;\pageref{fig:dendro}).  There are 171 sources with more than 95\,\% of their citation links in $\mathrm{\mathbf{L}}$, 160   of them are also in seed 298 (91\,\% of 175).

Clusters $\mathrm{\mathbf{TL}}$ and $\mathrm{\mathbf{TR}}$ are not valid on final resolution level but for $r=1/10$ and $r=1/4$,  respectively. For next levels, PsiMinL found a path through the cost landscape that ends in clusters $\mathrm{\mathbf{TLC}}$ and $\mathrm{\mathbf{R}}$,  respectively. All other clusters on intermediary levels are not considered here. They do not differ much from final clusters or are valid for  $r=1/20$ only.

%%%%%%%%%%
%  figure cost size
%%%%%%%%%%%%%
\begin{figure}[t]
\begin{center}
\includegraphics[height=6.1cm]{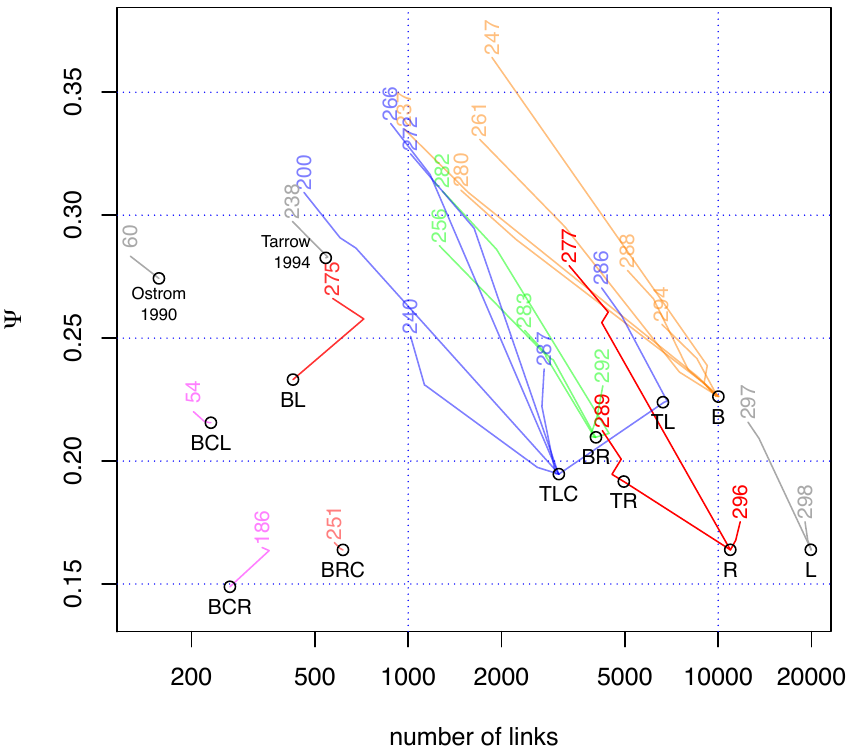}\caption{Cost-size diagram of 27 seed subgraphs and of steps towards 11 clusters valid on all resolution levels $r\le1/3$}
\label{fig:cost.size.27}
\end{center}
\end{figure}
%%%%%%%%%%%%%

%%%%%%%%%%
% table valid Ls
%%%%%%%%%
\begin{table}[t]
\caption{Link clusters ordered by seeds: first part -- Ward clusters of views, second part -- unions of valid clusters, third part -- CPLC-towns  (cf.\;text)}
\begin{footnotesize}
\begin{center}
\begin{tabular}{rlrr}
&name of	& number&  		\\
seed& cluster & of links &  $\Psi$ \\
\hline
54 & $\mathrm{\mathbf{ BCL }}$ & 231 & 0.21559\\
60 & Ostrom 1990 & 157 & 0.27430\\
186 & $\mathrm{\mathbf{ BCR }}$ & 266 & 0.14886\\
238 & Tarrow 1994 & 542 & 0.28265\\
251 & $\mathrm{\mathbf{ BRC }}$ & 616 & 0.16387\\
275 & $\mathrm{\mathbf{ BL }}$ & 425 & 0.23313\\
286 & $\mathrm{\mathbf{ TL }}$ & 6,634 & 0.22394\\
287 & $\mathrm{\mathbf{ TLC }}$ & 3,062 & 0.19469\\
289 & $\mathrm{\mathbf{ TR }}$ & 4,961 & 0.19167\\
292 & $\mathrm{\mathbf{ BR }}$ & 4,034 & 0.20966\\
294 & $\mathrm{\mathbf{ B }}$ & 10,015 & 0.22622\\
296 & $\mathrm{\mathbf{ R }}$ & 10,940 & 0.16392\\
298 & $\mathrm{\mathbf{ L }}$ & 19,895 & 0.16392\\
\hline
 $\mathrm{\mathbf{BCL}} \cup \mathrm{\mathbf{BCR}}$    &  $\mathrm{\mathbf{BC}}$ & 501 & 0.17894\\
 $\mathrm{\mathbf{BCR}} \cup \mathrm{\mathbf{BRC}}$   &$\mathrm{\mathbf{BRB}}$ & 893 &0.15913\\
 $\mathrm{\mathbf{TLC}} \cup \mathrm{\mathbf{TR}}$	&$\mathrm{\mathbf{T}}$	  &8,027 & 0.20321\\
\hline
   & Cox 1981 			& 536 & 0.27489\\
   &North 1990 				&265 &0.33104\\
   &Olson 1965			& 304 &0.35011\\
\end{tabular}
\end{center}
\label{tab.clusters}
\end{footnotesize}
\end{table}
%%%%%%%%%%%%

The first part of Table \ref{tab.clusters} lists data of all 13 clusters that have been reached from any of the selected 27 seeds. For clusters reached from more than one seed,  the first column gives the id number of the seed that is nearest in size to the final cluster.    
In some cases, different evolutions ended up in slightly different variants of a cluster. The best one invalidates the other variants.

According to the definition  in Equation \ref{eq.Psi} the cost function is equal for a link set and for its complement. Therefore, each complement of a  cluster  is also  a  cluster if it is a connected subgraph. Indeed, the largest valid cluster (with more than half of all links) is the complement of the second largest one: $\mathrm{\mathbf{L}} = E-\mathrm{\mathbf{R}}$.       

Complements of small subgraphs are nearly as large as the whole network and therefore not really interpretable as topics.
We therefore only consider the complement of cluster  $\mathrm{\mathbf{B}}$ (on size rank 3, with about one third of all $m=|E|= 30,835$ citation links). $E-\mathrm{\mathbf{B}}$ is connected but we have to test whether it  survives a local and a memetic search. That means, we have to use it as a seed subgraph for PsiMinL. $E-\mathrm{\mathbf{B}}$ remained unchanged and therefore valid till resolution level   $r=1/5$. On level $r=1/4$, PsiMinL invalidated $E-\mathrm{\mathbf{B}}$: it found a never rising path (with tunnels) through the cost landscape ending in  $\mathrm{\mathbf{R}}$. 

\begin{figure}[h]
\begin{center}
\includegraphics[height=10.7cm]{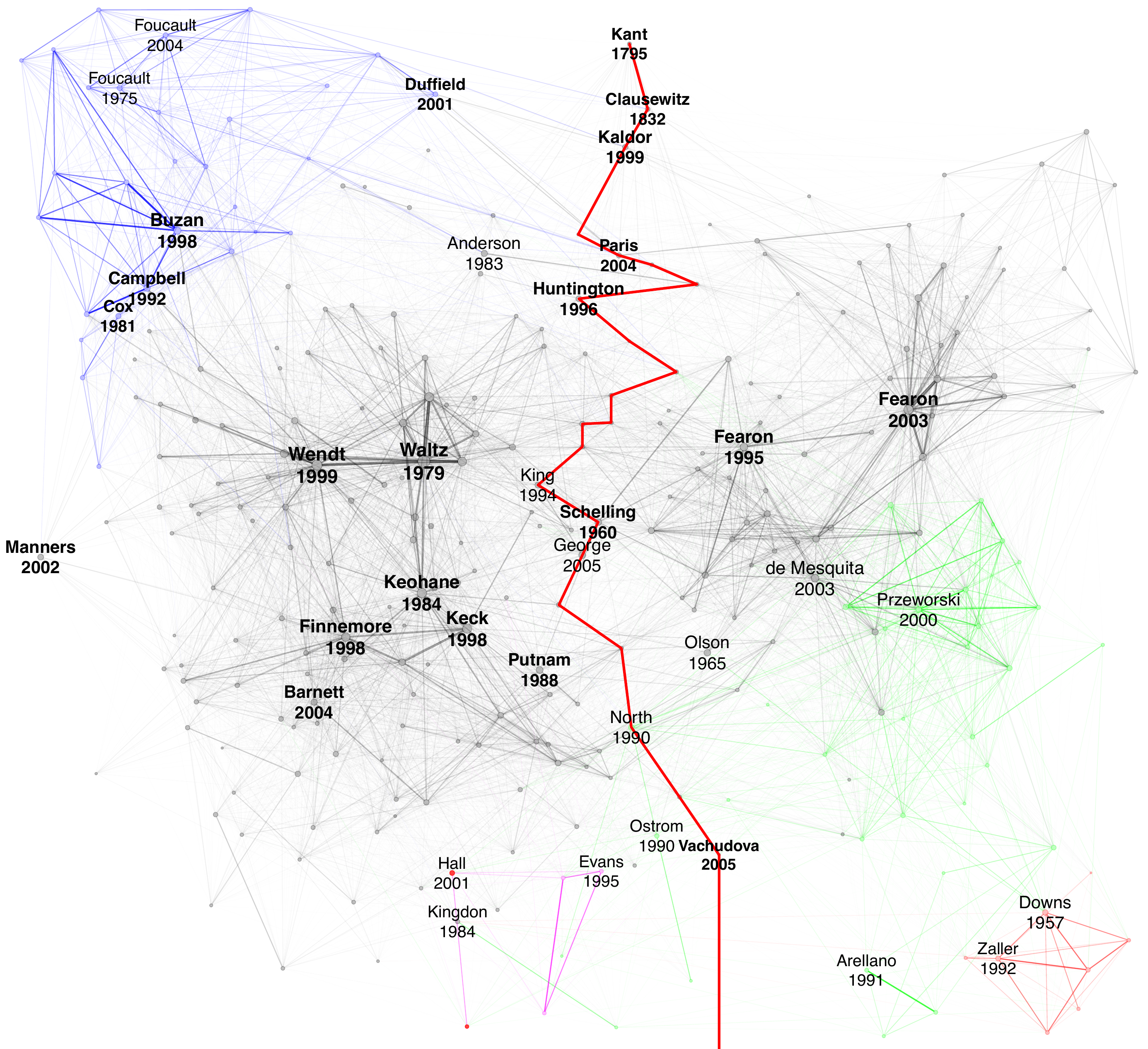}
 \includegraphics[height=10.7cm]{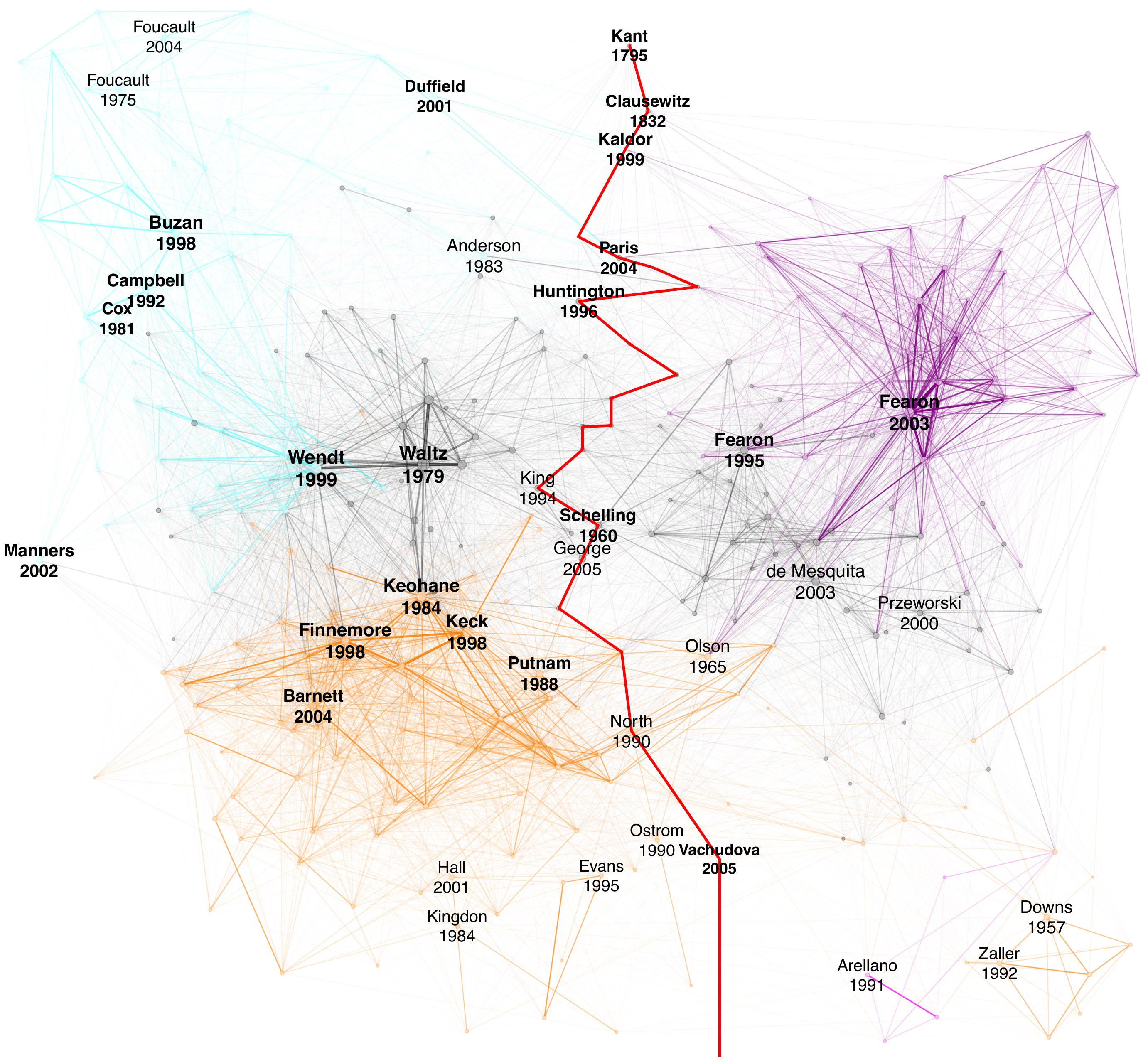}
 \caption{Graphs visualising valid clusters 2011--2015.  The red line marks the boundary between largest cluster $\mathrm{\mathbf{L}}$ on the left and its complement, the second largest cluster $\mathrm{\mathbf{R}}$ on the right. On the upper graph also clusters $\mathrm{\mathbf{TLC}}$, $\mathrm{\mathbf{BR}}$, $\mathrm{\mathbf{BL}}$, $\mathrm{\mathbf{BCL}}$, and $\mathrm{\mathbf{BRC}}$ are visualised, on the lower graph clusters $\mathrm{\mathbf{TL}}$, $\mathrm{\mathbf{TR}}$, $\mathrm{\mathbf{B}}$, and $\mathrm{\mathbf{BCR}}$ (cf.\;text).}
\label{fig:graphs}
\end{center}
\end{figure}

The bipartite network of papers and sources is very large. Therefore, clusters are visualised on a  projection of the bipartite network onto the co-citation graph of top-300 sources (Figure\;\ref{fig:graphs}). This has the wanted side-effect that a visual comparison of the two approaches can be made (s.a.\;footnote \ref{fn.bip}). We expect that link-cluster boundaries prefer regions of sparse co-citation relations. Following \citeN{marshakova1973ssm} edges between the 300 selected sources were weighted with their co-citation numbers diminished by expectation values derived from a null model of independent citations. Only edges that are significant on a 95\,\% level have been used as input for the force directed placement of nodes.\footnote{Fruchterman-Reingold algorithm, implemented in R-package \emph{sna} \cite{sna2016}}

The red line in the graphs of Figure \ref{fig:graphs} marks the boundary between $\mathrm{\mathbf{R}}$ on the right side and its complement $\mathrm{\mathbf{L}}$ on the left side of the graph. It connects  22 bridging sources that are cited by papers on both sides, beginning with Kant and von Clausewitz on the top and ending with Vachudova. More specifically, each of the bridging sources has  not less than 5\,\% of its citation links in each of the two complementary link clusters. All other sources have more than 95\,\% in $\mathrm{\mathbf{L}}$ or in $\mathrm{\mathbf{R}}$, respectively.

%%%%%%%%%%%%%%%%%%%%%%%%%
% figures: two graphs
%%%%%%%%%%%%%%%%%%%%%%%%%
\onecolumn
\twocolumn
%%%%%%%%%%%%%%%%%%%%%%%%%

Labels of sources are displayed for centres of 31 core-periphery structures obtained by running CPLC on the whole bipartite network and using results from the resolution level where the two most cited sources (Waltz 1979, Wendt 1999)  become independent from each other.  
I have added  labels of three sources at the ends of the red line (mentioned above) and of three sources that are centres in clusters (Arellano 1991, Evans 1995, Przeworski 2000). Labels are highlighted in bold for cited sources classified as belonging to the IR specialty.

A cluster is marked by colouring sources that have more than 95\,\% of their citation links inside its link set. A co-citation edge is coloured if more than half of all its co-citing papers have citation links (to the two sources) that belong to the cluster's link set. 
The colour used in Figure\;\ref{fig:cost.size.27} for a cluster is the same as in the graphs.\footnote{Citation numbers of the set of 300 selected sources restricted to citation links in valid clusters can be downloaded as R-object ccs-v7.RObj from  \url{https://zenodo.org/record/4181930} \cite{havemann_frank_2020_4181930}. File read-me.R contains R-code for listing core sources of clusters. The dataset on Zenodo also includes lists of sources in clusters and on their boundaries (file Havemann2020topics.pdf) and lists of journals with numbers of papers citing sources in clusters (file citing.journals.of.clusters.pdf).} 

Bold cluster names are derived from the position in the graphs of Figure \ref{fig:graphs}: $\mathrm{\mathbf{L}}$ -- left, $\mathrm{\mathbf{R}}$ -- right, $\mathrm{\mathbf{B}}$ -- bottom (orange),  $\mathrm{\mathbf{TR}}$ -- top right  (violet), $\mathrm{\mathbf{TL}}$ -- top left  (turquoise), $\mathrm{\mathbf{TLC}}$ -- top left corner (blue).\label{cluster.names} 
In the upper graph in Figure \ref{fig:graphs},  the red links and nodes represent cluster $\mathrm{\mathbf{BRC}}$ (bottom right corner). 

Pink elements correspond to cluster $\mathrm{\mathbf{BCL}}$ (bottom centre left). 
Cluster $\mathrm{\mathbf{BCL}}$ is also a subgraph of cluster $\mathrm{\mathbf{BL}}$ (bottom left, pink and dark red).
All these small clusters  are subgraphs of $\mathrm{\mathbf{BR}}$, which therefore is visualised not only by green nodes and co-citation links but includes all coloured elements in the bottom right of this graph.  

There are two small clusters in the first part of Table \ref{tab.clusters} that are  named after their most cited source, both with relatively high $\Psi$-values:

Cluster ``Tarrow 1994'' includes  
Sidney G. Tarrow's book \emph{Power in Movement: Social Movements and Contentious Politics} and five other sources with related themes, all outside IR and inside cluster $\mathrm{\mathbf{TR}}$. 

The cluster ``Ostrom 1990'' contains two sources with all their citation links:  Elinor Ostrom's  famous book  \emph{Governing the commons} (90 citations)  is  co-cited in 21 papers with \emph{The Tragedy of the Commons}, the paper by Hardin Garrett published 1968 in \emph{Science} (37 citations). 22 other sources have citation links within this cluster but get less than five citations from 106 papers belonging to it. The node with label ``Ostrom 1990'' can be found in the upper graph of Figure \ref{fig:graphs} near cluster $\mathrm{\mathbf{BL}}$ (bottom left, pink and dark red).

The second part of Table \ref{tab.clusters} lists data of new clusters reached by starting PsiMinL from seeds that are unions of  valid clusters in the first part. Unconnected unions cannot be seeds. 

The three smallest clusters and $\mathrm{\mathbf{BRC}}$ do not overlap each other in citation links, but one methodological book (Wooldridge 2002) is cited in all four clusters (by 30 papers in $\mathrm{\mathbf{BCR}}$, by one paper in each of the other three clusters). Thus, any union of them is a connected subgraph and can be used as a seed. 

Seeds made from unions of cluster ``Ostrom 1990'' with each of the other three small clusters did not bring any new result. In all three cases, cluster ``Ostrom 1990'' was excluded already on the first resolution level ($r=1/20$) and the other cluster was reached again by memetic search.

The union of $\mathrm{\mathbf{BCL}}$ and $\mathrm{\mathbf{BCR}}$  has  497 citation links and  $\Psi(\mathrm{\mathbf{BCL}} \cup \mathrm{\mathbf{BCR}})\approx 0.18079$. PsiMinL found the slightly better cluster $\mathrm{\mathbf{BC}}$ (bottom centre) on a short path through the cost landscape and already on resolution level $r=1/20$. All these statements hold analogously for cluster $\mathrm{\mathbf{BRB}}$ (bottom right bottom) which is not far from the union of $\mathrm{\mathbf{BCR}}$ and $\mathrm{\mathbf{BRC}}$.   Starting PsiMinL from $\mathrm{\mathbf{BCL}}\cup\mathrm{\mathbf{BRC}}$ ended up in cluster  $\mathrm{\mathbf{BRC}}$ itself already at the first resolution level.

Both new clusters, $\mathrm{\mathbf{BC}}$  and $\mathrm{\mathbf{BRB}}$, do not differ much from their seeds, which are (connected) unions of disjoint link sets. Thus, we can assume that they can easily be split into well separated parts. Indeed, running CPLC on, e.g., $\mathrm{\mathbf{BC}}$ results in two towns very similar to  $\mathrm{\mathbf{BCL}}$ and $\mathrm{\mathbf{BCR}}$, respectively,  already on resolution level $q=0$. Therefore, we can expect that  clusters $\mathrm{\mathbf{BC}}$  and $\mathrm{\mathbf{BRB}}$ are thematically not very homogenous. This can also be said about a cluster obtained from the union of cluster ``Tarrow 1994'' with cluster  $\mathrm{\mathbf{BCL}}$ (678 links, $\Psi \approx 0.25973$). 

Clusters $\mathrm{\mathbf{TLC}}$  and $\mathrm{\mathbf{TR}}$ overlap in only 24 links. Their union used as seed resulted in a new cluster with 8,027 links ($\mathrm{\mathbf{T}}$, for top), which is  valid on all levels.  96\,\% of all links in $\mathrm{\mathbf{TLC}}$ and 89\,\% of all links in $\mathrm{\mathbf{TR}}$ are also in  $\mathrm{\mathbf{T}}$. 

I conclude that seeds that are connected unions of disjoint (or nearly disjoint) link sets are not useful for identifying homogenous topics. 

Other (nontrivial) unions of overlapping clusters did not result in any new valid cluster. The same holds for intersections of valid clusters. 
Starting PsiMinL from  intersection $\mathrm{\mathbf{BR}} \cap \mathrm{\mathbf{L}} $, e.g., ended up with $\mathrm{\mathbf{BC}}$. 
I did not consider intersections of valid clusters that contain only a few links or more than 70\,\% of the links of the smaller cluster because  one can then expect that PsiMinL only finds this smaller cluster again. 

\begin{figure}[p]
\begin{center}
 \includegraphics[height=5.2cm]{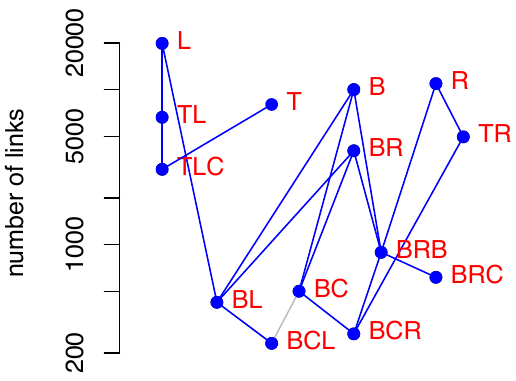}
  \includegraphics[height=4.7cm]{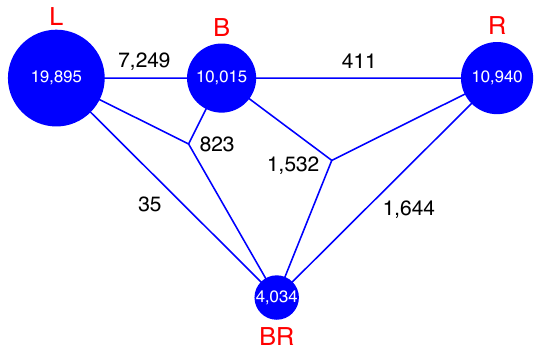}
 \caption{Left: poly-hierarchy of clusters (note log-scale of size); right: overlaps of  four clusters without super-clusters  (size of triple overlap subtracted from size of pairwise overlaps, cf.\;Table \ref{tab.overlap})}
\label{fig:poly-hierarchy}
\end{center}
\end{figure}

The left-hand side of Figure \ref{fig:poly-hierarchy} visualises the poly-hierarchy of clusters. A blue line is drawn if the smaller cluster has less than 5\,\% of its links outside  the larger cluster.

The tiny cluster $\mathrm{\mathbf{BCL}}$ 
has 215 of its 231 links (93.1\,\%) in $\mathrm{\mathbf{BC}}$ and is totally included into $\mathrm{\mathbf{BL}}$. 
Total inclusion is the exception. This is due to the normalisation in Equation \ref{eq.Psi}. The cost of the smaller cluster is lower with some additional links but not the cost of the larger cluster because a smaller link set has a larger relative increase of the denominator $k_\mathrm{in}(L)$ by including links than a larger set.

On the right-hand side of Figure \ref{fig:poly-hierarchy}, overlaps between four clusters are displayed that are not (nearly totally) included in a larger cluster.   $\mathrm{\mathbf{L}}$ and $\mathrm{\mathbf{R}}$ have zero overlap by definition. 823 of all 858 links in $\mathrm{\mathbf{L}} \cap \mathrm{\mathbf{BR}}$ are also in $\mathrm{\mathbf{B}}$. The remaining 35 citation links are visualised by the direct edge between $\mathrm{\mathbf{L}}$ and $\mathrm{\mathbf{BR}}$. 
The edge betwen $\mathrm{\mathbf{B}}$ and $\mathrm{\mathbf{BR}}$ is missing because all 2,345
links in $\mathrm{\mathbf{B}} \cap \mathrm{\mathbf{BR}}$ are either in $\mathrm{\mathbf{L}}$ or in its complement $\mathrm{\mathbf{R}}$ (s.\;Table \ref{tab.overlap}).

%%%%%%%%%%%%%%%%%%%%%%%%%%%
% Table Overlaps...
%%%%%%%%%%%%%%%%%%%%%%%%%%%
\begin{table}[t]
\caption{Overlaps of four valid clusters without super-clusters (cf.\;text and Figure \ref{fig:poly-hierarchy})}
\begin{center}
\begin{footnotesize}
\begin{tabular}{rr}
link set &  links \\ 
\hline
$\mathrm{\mathbf{L}} \cap \mathrm{\mathbf{B}}\cap \mathrm{\mathbf{BR}} $& 823 \\\
$\mathrm{\mathbf{R}} \cap \mathrm{\mathbf{B}}\cap \mathrm{\mathbf{BR}} $& 1,532\\\
$\mathrm{\mathbf{L}} \cap \mathrm{\mathbf{B}}$& 8,072\\
$\mathrm{\mathbf{L}} \cap \mathrm{\mathbf{BR}}$& 858\\
$\mathrm{\mathbf{R}} \cap \mathrm{\mathbf{B}}$& 1,943\\\
$\mathrm{\mathbf{R}} \cap \mathrm{\mathbf{BR}}$& 3,176\\
$\mathrm{\mathbf{B}} \cap \mathrm{\mathbf{BR}}$& 2,355 
\end{tabular}
\end{footnotesize}
\end{center}
\label{tab.overlap}
\end{table}%
%%%%%%%%%%%%%%%%%%%%%

%%%%%%%%%%%%%%%%%%%%%%%%%
\subsection{Core-periphery structures}
\label{sec.cp}
%%%%%%%%%%%%%%%%%%%%%%%%%

Constructing core-periphery structures of a cluster can reveal its highly cohesive cores if it has  one or more of such cores. 
Clusters in the second part of Table \ref{tab.clusters} decay into two  well separated sub-clusters.  We can therefore neglect them when we look for cohesive cores.

\begin{figure}[p]
\begin{center}
 \includegraphics[height=10.2cm]{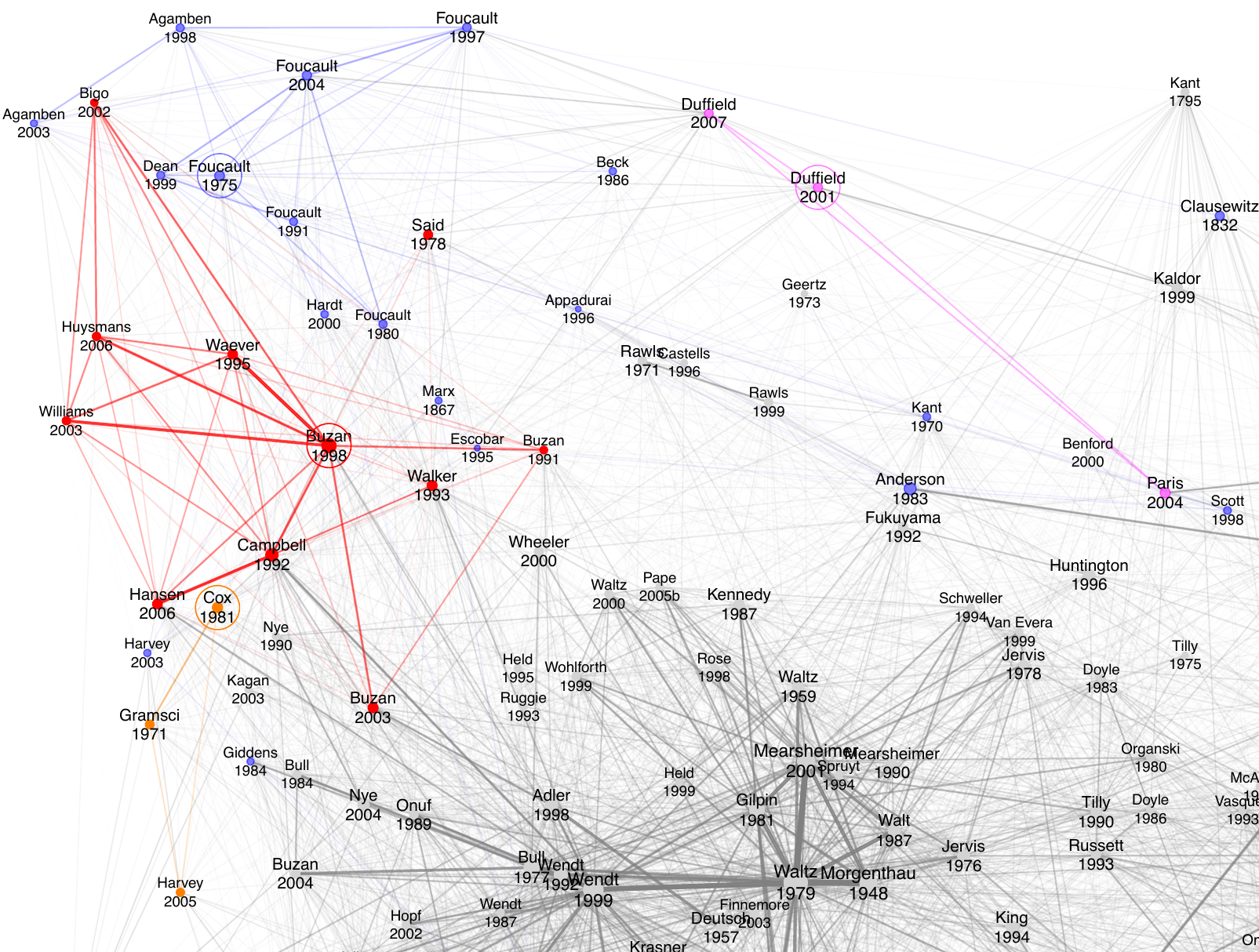}
 \caption{Four core-periphery structures (towns) in cluster $\mathrm{\mathbf{TLC}}$ visualised by different colours. Central sources are marked by a circle.}
\label{fig:CPLC}
\end{center}
\end{figure}

For all other 11 valid clusters found on resolution level $r=1/3$, core-periphery structures (towns) were constructed by running CPLC for a sequence of values of resolution parameter $q \in [0, 1/2]$.  
Figure \ref{fig:CPLC} shows the four towns in $\mathrm{\mathbf{TLC}}$ obtained by CPLC on resolution level $q=0.183$. The pale blue town around Foucault (1975) has a larger periphery than the three other towns. I here only present this example, which at least gives cursory evidence that CPLC indeed reveals core-periphery structures in clusters. I have to leave a detailed examination of results  to further work. 

Towns of clusters were also used as seed subgraphs for finding further clusters. One example is a town of $\mathrm{\mathbf{L}}$ with Wendt (1995) as the centre. Starting from this seed, PsiMinL rediscovered cluster $\mathrm{\mathbf{TL}}$. I selected those towns as seeds which promised to lead to new clusters from inspecting the co-citation graph (Figure \ref{fig:graphs}).
Further successful cases are the three clusters in the third part of Table \ref{tab.clusters},  which are named after the centres of their seed towns. 

The paper by Robert Cox (1981) about \emph{Social Forces, States and World Orders} can be found on the left-hand side of  graphs in Figures  \ref{fig:graphs} and \ref{fig:CPLC}.  It is often co-cited with Marx and Gramsci and with two books by David Harvey published 2003 and 2005, respectively.  These five sources are the sources with full membership in this cluster and also with all their citation links inside cluster $\mathrm{\mathbf{TLC}}$.

The book by Douglass North (1990, on the red line in Figure  \ref{fig:graphs}) is significantly often co-cited with the book by Oliver Williamson (1985), both dealing with economic institutions. They have all their citation links in this cluster. The next relevant source is  Ostrom's book (1990), which is cited by ten cluster papers but gets 90 citations in the whole set.  

Mancur Olson's book about \emph{The Logic of Collective Action} (1965, on the right side of the red line in graphs of Figure \ref{fig:graphs}) is the only full-member source in its cluster. In contrast to the other two clusters in the third part of Table  \ref{tab.clusters}, this cluster remains valid only till $r = 1/5$. For $r=1/4$, PsMinL invalidated it by reaching $\mathrm{\mathbf{BCR}}$.

%%%%%%%%%%%%%%%%%%%%%%%%
% figures
%%%%%%%%%%%%%%%%%%%%%%%%
\onecolumn
\twocolumn
%%%%%%%%%%%%%%%%%%%%%%%%%%%%%%%%

%%%%%%%%%%%%
\section{Discussion}
%%%%%%%%%%%

\subsection{Clustering method} 

Methods for the clustering of networks can use global evaluation functions that evaluate whole partitions, like modularity, or local functions that evaluate each cluster independently from others, 
like conductance or normalised cut  for node clustering \cite{fortunato2010community} and normalised node-cut $\Psi$ for link clustering.

Topics are locally defined. This favours the use of local evaluation functions  for topic reconstruction. Citation links are the thematically least heterogeneous bibliometric elements. This suggests to apply link clustering algorithms  in citation networks.  Topics can overlap and form a poly-hierarchy, which in turn means that topic clusters should not be too hard to split into sub-clusters. Thus cohesion cannot be the main criterion for evaluating a cluster. Until now PsiMinL is the only algorithm that is in line with all these demands. The price payed for this are long running times, the need for many CPUs,  and a high complexity of the whole analysis (s.a.\;the  discussion of computer running times of PsiMinL in Appendix on p.\;\pageref{sec:run.times}). 

Next to these abstract and technical considerations, the crucial test relates to domain knowledge: 
Can experts interpret not only single clusters but also the  poly-hierarchy they form and their overlaps?\footnote{Otherwise, all the effort becomes problematic. A further interesting question is, whether one finds top sources in overlaps that are cited for different reasons in different overlapping clusters, which was one of our arguments for clustering citation links.}   I have to leave this for further work. 

This paper makes several novel contributions. For the first time, I apply PsiMinL  to a bipartite network of highly cited sources and papers citing at least two of them. I argue that this restriction is possible because top-cited sources serve as symbols for shared knowledge of a scientific community in a field and shared knowledge is what a topic defines. This restriction reduces the network size (by a factor of ten) and therefore also the computational effort. Also for the first time, I overcome the somewhat arbitrary choice of a fixed resolution by going through a sequence of resolution levels and using the resulting clusters on one level as seeds for the next one. A further novelty is that I construct initial seed subgraphs from clusters corresponding to long branches in the dendrogram obtained by Ward co-citation clustering. This is also the first PsiMinL analysis of a specialty belonging to the social sciences.

\subsection{Clustering results} 

Three different data models were used here, namely:
\begin{enumerate}
\item
the bipartite network of top-300 sources and all papers in IR journals and books citing at least two of them (used by link clustering algorithm PsiMinL, leading to a poly-hie\-rarchy of clusters),
\item
the projection of the bipartite network onto the co-citation graph of top-300 sources (on which clusters are displayed after selecting significant links), and
\item
a distance matrix between top-300 sources made from the co-citation projection weigh\-ted with Salton's cosine (used for constructing seed subgraphs from Ward clusters of views).
\end{enumerate}

In spite of data differences, each link cluster concentrates in a certain region of the co-citation graph. Most of clusters have boundaries going to sparse regions of the graph. 
This is a first hint that PsiMinL applied on a bipartite network of papers and top-cited sources  leads to reasonable clusters. I have to leave any further evaluation of contents of PsiMinL clusters and of their core-periphery structures obtained here to IR experts \shortcite{Risse2020IR}.

I can, however, compare these clusters quantitatively with all clusters of views on all levels of hierarchical Ward clustering. How many top-300 sources of a Ward cluster are core members of any link cluster? 
The results are presented in Appendix (p.\;\pageref{sec:comp.Ward.Psi}). 

Three link clusters are never a best match of a seed, namely those made from unions of two clusters: $\mathrm{\mathbf{BC}}$, $\mathrm{\mathbf{BRB}}$, and $\mathrm{\mathbf{T}}$ (second part of Table \ref{tab.clusters}, p.\;\pageref{tab.clusters}). This corresponds to their probable thematic inhomogeneity discussed above.  

There are five exact matches between clusters, which all have less than seven cited sources (Table \ref{tab.match}, p.\;\pageref{tab.match}). The worst match is with cluster $\mathrm{\mathbf{TL}}$ (Salton's cosine $s\approx0.76$). The   division between the two largest clusters $\mathrm{\mathbf{L}}$ and $\mathrm{\mathbf{R}}$ is matched with values of $s>0.9$.

All but one of the matched link clusters in the first part of Table \ref{tab.clusters} are matched best by their (nearest) seed. Only  $\mathrm{\mathbf{TLC}}$  is best matched by a Ward cluster that is not in the set of 27 long-branch seeds but among the 23 seeds with shorter branches (cf.\;footnote \ref{shorter}).   PsiMinL reaches $\mathrm{\mathbf{TLC}}$ from this seed too. 

How can we interpret these good matches between link clusters and some Ward clusters of views that correspond to long branches in the dendrogram? 

First, the  two approaches are compatible and therefore supporting  one another. 

Second, the use of long-branch clusters as seed subgraphs for PsiMinL is confirmed as an efficient method. Starting from seeds from a global cut through the dendrogram needs longer paths in the cost landscape and resulted only in a subset of valid link clusters obtained with long-branch seeds.  That means, starting from long-branch seeds we rediscover all clusters that were found with global-cut seeds. In other words,  
similarity of seeds and resulting clusters is not the reason for finding this set of clusters.

Experiments with seeds corresponding to 23 branches with sub-maximal length in their size classes showed that we can find  more small valid link clusters when starting from small seeds with shorter branches too (cf.\;Appendix, p.\;\pageref{23-seeds}). 
Some of these small clusters  are not as well separated as the best clusters in Table \ref{tab.clusters} (p.\;\pageref{tab.clusters}). Their  $\Psi$-values exceed 1/4 (cf.\;also Table  \ref{tab.7.clusters}, p.\;\pageref{tab.7.clusters}).  

Evaluation function $\Psi$ is always larger than the escape probability of the random link-node-link walker \cite{evans2009line}, for small clusters  only slightly larger, because the denominator of the second term in the definition of $\Psi$ (Equation\,\,\ref{eq.Psi}) is very large. That means, for $\Psi < 1/2$  the random walker's probability to remain within the cluster  is always larger than to escape from it in the next step ($P_\mathrm{esc} < 1/2$). 

An ordinary random walker hopping from node to node escapes from a \emph{weak} node community as defined by \shortciteN{Radicchi2004defining} also with a probability $P_\mathrm{esc} < 1/2$.   
Translating the definition of weak communities  into the language of  link clustering \shortcite{havemann_communities_2019}, we can deduce that all clusters obtained here are link communities in the weak sense.  

Recently  \citeN{Kristensen2018IR-end}   determined disjoint  co-citation clusters of 332  authors highly cited as first authors in 106 IR journals in the period 2011--2015. His aim was to visualise the ``communicative-sociological structures'' of the discipline. He admits that neglecting co-authors of highly cited first authors can cause biases towards some authors, especially towards authors of theorising works. He found some authors with a ``fairly stable position in the network'' but others ``whose work is used for positioning by several camps may shift camps depending on the specific threshold values" (p.\;247). 

In my approach each highly cited work can appear in more than one cluster because I produce overlapping clusters of cited sources. Topics overlap in authors even more than in papers or books but at first glance both networks show at least some similar structures. 
The contents of  Kristensen's \textit{camps} of authors and of link clusters obtained here cannot be compared without knowledge of the field.

%%%%%%%%%%%%%%%%%%%%%
\section{Conclusions}
%%%%%%%%%%%%%%%%%%%%%

Can PsiMinL be recommended for finding a poly-hierarchy of overlapping research topics of a specialty? 
The experiments made in this study suggest that we indeed obtain reasonable results by applying PsiMinL to a bipartite network of selected concept symbols and all papers citing at least two of them.\footnote{One caveat  has to be made: Researchers in International Relations as in other specialties in social sciences often refer to books as concept symbols. 175 of the top-300 sources are books (s.\;Table \ref{tab.top.300.distr}, p.\;\pageref{tab.top.300.distr}). Thus, a success of the approach for specialties of natural science can be expected but not guaranteed.}   
IR experts were able to interpret them \shortcite{Risse2020IR}. All resulting clusters were only slightly changed after adding missing links to the network  (s.\;Appendix, p.\;\pageref{sec.comment}). Several link clusters have a good match with Ward clusters of views (Table \ref{tab.match}, p.\;\pageref{tab.match}). A comparison with results of further  clustering algorithms applied on the same data would be useful for evaluating the new approach to clustering concept symbols. A first trial with classic co-citation analysis (single linkage of cosine weighted links) as done by \citeN{small1985ctc} was made. Also here,  results suffer from chaining, the well-known disadvantage of  single linkage. 
Differences between clusters obtained by PsiMinL and by other algorithms could by evaluated by experts of the specialty. I have to leave such comparisons to further work. 

Generally, any partition of a network into disjoint clusters cannot be compared as a whole with a poly-hierarchy of overlapping clusters. A good matching of all clusters is only possible, if the  clusters used for a quantitative comparison form  a hierarchy that has many levels (like the Ward cluster of views discussed above). 

Similar results  of different clustering methods can be seen as a mutual support but different results do not falsify any of the methods. They can be interpreted as reconstructing legitimate alternative  perspectives on the structure of a specialty's literature \cite{glaser2017same}. At most, one method could  be judged as more accurate than the other when we compare both with regard to the purpose of clustering  \cite{waltman2020principled}. A poly-hierarchy of independently evaluated clusters, as delivered by PsiMinL, could represent already different perspectives on the analysed literature. 

Evaluation function $\Psi$ can be justified within the model of a random walker who should leave a cluster with low probability  \shortcite{havemann_communities_2019}. For finding node clusters, each step of a random walker starts and ends on a node. Link clusters can be constructed by starting and ending on links  \shortcite{evans2009line}.\footnote{Recently, a random link-node-link walker's escape probability was used by \shortciteN{Enders-doi:10.1111/geb.13082} to cluster 39 standard hypotheses about biological invasions for mapping this specialty. }
 Random walks last long in well separated clusters. When a cluster contains sub-clusters which are only weakly connected with one another the chance to leave it can nonetheless be as low as to leave any of the two sub-clusters. In this sense random walkers are insensitive to inner cohesion of clusters. I argue that we need cohesion insensitivity when we want to obtain hierarchically organised sets of clusters. Only the smallest clusters can be expected not to decay into sub-clusters.

Seeing a research topic as a shared focus on scientific knowledge suggests that not separation but cohesion of views on knowledge should be the defining property of topics. We have tried to weaken this argument by pointing to core-periphery structures and by proposing the simple CPLC algorithm that constructs such structures inside well separated link clusters \shortcite{havemann_communities_2019}. This approach still rests on the assumption that topics can be represented by well separated clusters. The experiments with PsiMinL show that there are such topics but they do not prove that all research topics can be separated from the rest of a citation network. In dense cores of the network, separation could fail as the occurrence of a \emph{terra incognita} (a huge central cluster without substructures) in the analysis of astronomy and astrophysics seems to suggest \shortcite[p.\;1105]{havemann_memetic_2017}. 

Technically, PsiMinL is an evolutionary algorithm that searches for local minima in the cost landscape with evaluation function $\Psi$. 
PsiMinL starts memetic evolutions from seed subgraphs but  the same valid cluster can be reached from different seeds (cf.\;Figure \ref{fig:cost.size.27}, p.\;\pageref{fig:cost.size.27}). In this sense, the cluster solution is independent from seeds.  The construction of seeds influences only the time needed for a solution and its completeness.
 
All technical parameters of PsiMinL also do not affect the results but only the time needed to obtain them. The only numerical parameter that influences the shape of clusters is resolution $r$. In this study, I have tested a procedure that makes the results less dependent on $r$. I started with low $r$ and then iteratively used the clusters as seed subgraphs for running PsiMinL for higher levels of $r$. Because lower $r$ means faster search this strategy could also be advantageous when  results on only one resolution level are needed. 

Evolutionary algorithms on large networks need much computing time. PsiMinL as other algorithms shifts the time problem at least partly to one of computing power by applying highly parallel procedures. Genetic operators can be applied parallelly on all individuals of a population. Because clusters are evaluated independently we can start PsiMinL parallelly from different seeds. Further optimisation of PsiMinL could be reached by finding optimal sets of technical parameters like population size, mutation rate etc. Another technique for reducing computing time could be to start with only two years and then use resulting clusters as seeds for larger periods, similar to reducing a large graph by  random sampling   \cite[p.\;23]{azaouzi_community_2019}.

Finding a minimum in a large and rough cost landscape by applying an evolutionary strategy never comes to an end because we cannot prove that there is no lower place than the one found. PsiMinL searches for local minima and accepts a link cluster $L$ as a valid solution if it is not made invalid by a lower place inside a radius of $r|L|$ in the landscape. That means, here we cannot exclude that there are better variants of clusters but we also cannot maintain that we have found all valid clusters. Sometimes, PsiMinL invalidates a cluster not in the first trials. That means, we cannot be sure that a found cluster is really valid, but we can at least assume a weak validity when PsiMinL is not able to find a path to a better cluster after several trials.  

Applying PsiMinL for finding link clusters in citation networks needs preprocessing (data cleaning, 
construction of seeds)\footnote{But note, that tedious cleaning of citation data can be reduced to highly cited sources when citation networks of concept symbols are clustered.} 
and postprocessing (selection of valid solutions, finding cohesive cores). Running PsiMinL many times for many seeds requires not only computing time and power but also a clear organisation of all procedures, selections, and validations. PsiMinL cannot be recommended for a user only interested in results before the whole procedure is transformed into a routine of automatic actions. 
More experience is needed for optimising the exploration of cost landscapes with PsiMinL. Then, hopefully, we can make a step further in codifying the procedure.

%\newpage
%%%%%%%%%%%%%%%%%%%%%%%%%%%%%%%
\section*{Acknowledgements}
%%%%%%%%%%%%%%%%%%%%%%%%%%%%%%%
This work is part of the Global Pathways project sponsored by DFG (grant RI 798/11-1).\footnote{\url{http://t1p.de/globalpathways}} As a member of the project team, Felix Mattes, made all downloads and developed the algorithm for reference identification. Lixue Lin-Siedler helped in classifying references as scholarly ones. The team members and experts in international relations, Thomas Risse, Wiebke Wemheuer-Vogelaar, and Mathis Lohaus commented on results and classified the 300 highly cited sources. Special thanks to Jochen Gl{\"a}ser who gave valuable advices in the whole process of data collection and processing. 
The memetic algorithm was implemented as an R-package by Andreas Prescher in a project funded by the German Research Ministry (BMBF grant 01UZ0905). I thank Michael Heinz for many discussions and for applying an alternative clustering method. 
He, Jochen Gl{\"a}ser, Alexander Struck, Mathis Lohaus, and Martin Enders also commented on drafts of the paper.  The comments of two anonymous reviewers were also very helpful for improving the paper, many thanks!
Finally I thank the developers of  \LaTeX\,\,and of R.\footnote{\url{https://www.r-project.org}} 

\section*{Data availability}
The raw data used in this paper were obtained from the Web of Science database produced by Clarivate Analytics. Due to license restrictions, the data cannot be made openly available. To obtain Web of Science data, please contact Clarivate Analytics.\footnote{\url{https://clarivate.com/products/web-of-science}}

Results of cleaning and clustering can be found on Zenodo \cite{havemann_frank_2020_4181930}.
%\twocolumn

\section*{Competing interests}
The author declares that he has no competing interests.

\newpage
%%%%%%%%%%%%%%%%%%%%%%%%%%%%%%%%%%%%%%%%%%%%%%%%%%%%%%%%%%
\section*{Appendix}
%%%%%%%%%%%%%%%%%%%%%%%%%%%%%%%%%%%%%%%%%%%%%%%%%%%%%%%%%%
\label{app}
\subsection*{Data}

In Table \ref{tab.journals} (p.\;\pageref{tab.journals}) all downloaded journals  are listed together with numbers of  papers with references in the whole 10-years period and in both 5-years periods, respectively.\footnote{The full names of journals can be read in the csv-file \emph{journals.csv} which can be downloaded from  \url{https://zenodo.org/record/4181930} \cite{havemann_frank_2020_4181930}.}

\label{data.exp}

We identified references to highly cited sources in the set of downloaded papers. Among these, our IR experts selected 137 IR books and 102 IR papers. The paper set was expanded by downloading 14,389 papers citing these sources but published in 3,135 non-IR journals and serials and in 2,405 non-IR books.\footnote{Data expansion to papers in non-IR journals and books was neglected for the clustering exercise because it would bias the results towards the views of authors outside IR.\label{fn.exp} These papers have been downloaded to get a comprehensive bibliography of IR papers.}

All records without references were excluded  because they would not be part of a citation-based network. This results in a total number of 71,210 records of the following types:
 
\begin{tabular}{rl}
53,889 & original papers, \\
15,524& book reviews, and\\
1,797 & review articles.\\
\end{tabular}

Because not all variants of references to one and the same source are identified in WoS we need some further identification, but we have excluded references to non-scholarly historical sources from this tedious task. 
Thus, among the reference strings of publications, a subset was categorised as probably referring to scholarly sources when they included an author name, a year, and did not refer to a newspaper. To these  references  an algorithm for reference identification was applied, which is described below.

 All in all, 1,143,317 different reference strings to scholarly sources were identified. They represent about two millions of citations of  992,582 identified sources. In many papers the authors refer to different pages of a source---whether be it a book or a paper. Because  such references were identified here (but generally not in WoS) there are some duplicated links between citing papers and cited sources.   
 24,308 citing papers are also cited sources.

\label{app:ref-id}
References where identified by an algorithm that Felix Mattes implemented as a script in R \cite{R2015}. It calculates distances between elements of different reference strings (names of authors  and of journals, book titles, volumes, pages, and DOIs). String distance is based on OSA distance (Optimal String Alignment, a restricted Damerau-Levenshtein distance) in R-package \emph{stringdist} \cite{vanderLoo2014stringdist}.  Distance is defined as zero when all characters of the shorter string occur in the same order in the longer one. Beginning pages are allowed to have a maximum difference of ten. Publication years must be equal. After calculating distances for each element of references, distances between whole references are determined as a weighted sum of element distances. We experimented with different weights and thresholds to obtain a criterion for identification which avoids false positives but also false negatives. 
The results of reference identification can be down\-loaded from \url{https://zenodo.org/record/4181930} (file \emph{ref.sou.csv}). 
 There you also find the R-script used for reference identification \cite{havemann_frank_2020_4181930}.
 
\label{top-300}
Table \ref{e.tab.1} on pp.\;\pageref{b.tab.1}--\pageref{e.tab.1} gives the top-300 sources highly cited in IR papers 2006--2015 in alphabetical order of the first author or editor (co-authors and co-editors can be found in the last column). 

The selection is based on citation numbers in the expanded data set. Therefore it is biased  towards the 239 IR sources used for data expansion (because highly cited non-IR sources could have less gain in citation numbers from data expansion). This bias has to be considered when results are interpreted.\footnote{The citation numbers in column \textit{cit} of Table \ref{e.tab.1} on pp.\;\pageref{b.tab.1}--\pageref{e.tab.1} equal the numbers of citing papers in IR journals only, i.e, without data expansion.\label{fn:exp} There is a large overlap between the 203 IR sources among the top-300 and the 239 IR sources used for data expansion.}

The column \textit{cit} shows the numbers of papers citing the sources in 2006--2015. The ranks in the first column are determined according to these citation numbers. Ties are broken by year of first publication, beginning with older sources. Ranks in bold numbers belong to sources classified as IR sources by experts. Table \ref{tab.top.300.distr} contains numbers of top-300 sources with regard to publication type and content.

%%%%%%%%%%%%%%%%%%%%%%%%%%%%%%%%
% Table books-papers, IR-non-IR
%%%%%%%%%%%%%%%%%%%%%%%%%%%%%%
\begin{table}[h]
\caption{Distributions of top-300 sources with regard to publication type and content }
\begin{center}
\begin{tabular}{r|rr|r}
	&	 IR & non-IR	&sum\\
	\hline
books &	106 &	69 &	175\\
papers &	 97	&	28 &	125\\
	\hline
sum	&	203	 &   97  & 300\\
\end{tabular}
\end{center}
\label{default}
\label{tab.top.300.distr}
\end{table}%
%%%%%%%%%%%%%%%%%%%%%%

The oldest top-300 source is Kant's  \textit{Perpetual Peace} (1795).
There are two further sources that  were first published before 1900 (von Clausewitz and Marx), and three in the first half of the twentieth century (Weber, Polanyi, and Morgenthau).  

\begin{figure}[t]
\begin{center}
 \includegraphics[height=4.3cm]{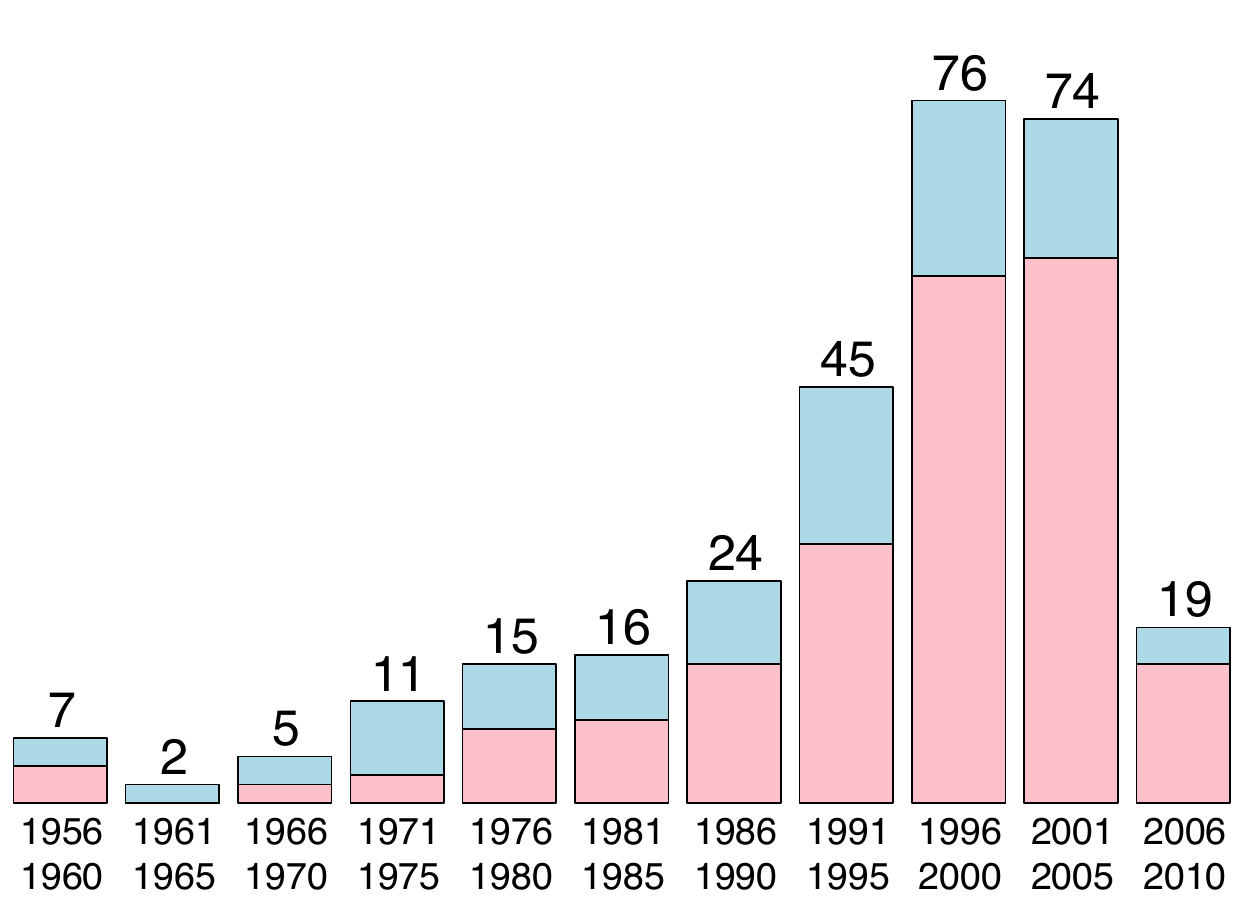}
 \caption{Distribution of first publication year of  highly cited sources (after 1950; 294 of 300 sources; 5-years periods). Red parts of bars represent IR-sources, blue parts sources outside IR.}
\label{fig:time.sou}
\end{center}
\end{figure}

In Figure \ref{fig:time.sou} the distribution of publication years of all other 294 highly cited sources is visualised.  In 2010 we have three sources (Blattman, Carter, and Cheibub). Three sources are also in the set of IR-papers downloaded from WoS (Powell  2006,  Hegre   2006, and Hooghe  2009).

%%%%%%%%%%%%%%%%%%%%%%%%%%%
\subsection*{PsiMinL glossary}
%%%%%%%%%%%%%%%%%%%%%%%%%%%%
\label{sec.gloss}
\begin{description}
\item[cost landscape] 
a set of places (link sets in a network, neighbouring places differ in one link) with a height (cost $\Psi$)

\item[crossover] 
combination of genes of two individuals, here: union or intersection of them

\item[individual]
connected set of links in a graph  

\item[initialisation] generating a population by mutating an individual several times followed by local searches until enough different individuals are produced

\item[genes]
the links  of an individual

\item[generation]
a round of memetic search including mutation, crossover, and selection

\item[memetic algorithms]
evolutionary algorithms that also include deterministic local sear\-ches in a cost landscape

\item[mutation]
random change of some genes of the best individual

\item[offspring]
new individuals made by crossover between the current best individual and most different other individuals in the population

\item[population]
a set of individuals

\item[renewal]
a new initialisation made after some generations without improvement

\item[selection] final genetic operator in each generation that selects for the new generation of a population the best individuals from the union of old population, its offspring, and its mutants

\item[tunneling]
a metaphor for going through barriers in the cost landscape (which is done in a local search if the tunnel is not too long, i.e., if the link set at its end invalidates the link set at its entry)    

\item[valid cluster] 
a cluster $L$ that differs in at least $r|L|$ links from any cluster with lower cost $\Psi$ ($0< r < 1$ measures resolution)

\end{description}

%%%%%%%%%%%%%%%%%%%%%%%
\subsection*{Selecting seeds}
%%%%%%%%%%%%%%%%%%%%%%%
\label{app:seeds}
\label{23-seeds}
\begin{figure}[p]
\begin{center}
 \includegraphics[height=20.7cm]{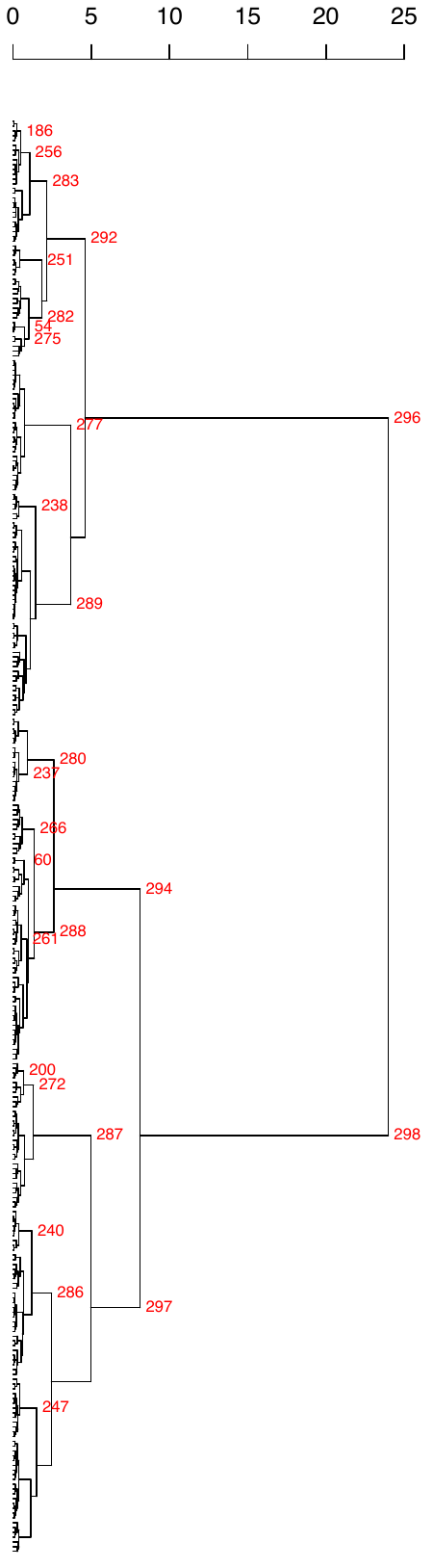}%
\caption{Dendrogram of Ward clustering of co-cited top-300 sources. Red numbers are identifiers of Ward clusters. The horizontal axis displays variance of clusters.}
\label{fig:dendro}
\end{center}
\end{figure}

\begin{figure}[!t]
\begin{center}
 \includegraphics[height=10cm]{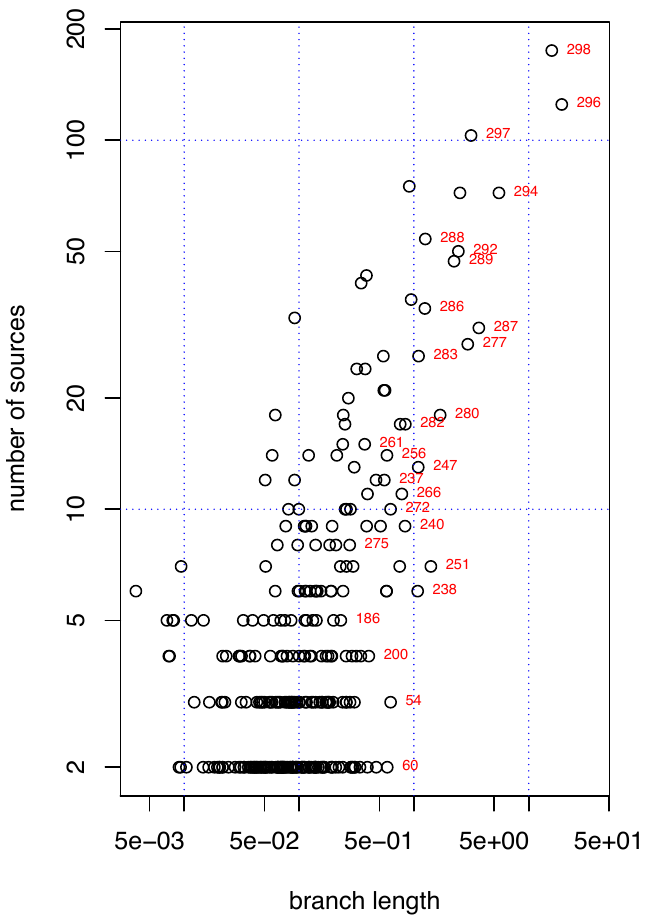}
 \caption{Branch lengths for all cluster sizes in the dendrogram of Ward clustering of views (log scales on both axes). Red numbers are the same identifiers of Ward clusters as in Figure \ref{fig:dendro}.}
\label{fig:branch.length}
\end{center}
\end{figure}

The dendrogram of Ward clustering of views is in Figure \ref{fig:dendro}.
In Figure \ref{fig:branch.length} branch lengths in the Ward dendrogram are displayed  for all cluster sizes. Red numbers in both figures are identifiers of 27 selected Ward clusters with longest branches in their size classes. They are used for constructing   seed subgraphs for memetic evolutions.

%%%%%%%%%%
% table new valid Ls
%%%%%%%%%
\begin{table}[!b]
\caption{Seven further link clusters from Ward clusters of views with shorter branches (cf.\;text)}\begin{center}
\begin{tabular}{rrr}
	& number&  		\\
seed & of links &  $\Psi$  \\
\hline
79 & 227 & 0.27092\\
82 & 176 & 0.29611\\
88 & 229 & 0.29123\\
196 & 542 & 0.24096\\
219 & 432 & 0.28063\\
221& 293 & 0.33220\\
249 & 150 & 0.32511\\
%\vspace{0.1cm}
\end{tabular}
\end{center}
\label{tab.7.clusters}
\end{table}
%%%%%%%%%%%%%%%

I have checked whether seeds made from further 23 Ward clusters with shorter branches in the dendrogram results in new valid clusters. Starting from 16 of them, PsiMinL rediscovered six clusters (five times $\mathrm{\mathbf{TLC}}$, three times $\mathrm{\mathbf{BR}}$ and $\mathrm{\mathbf{BCR}}$, two times $\mathrm{\mathbf{L}}$ and $\mathrm{\mathbf{R}}$, and one time $\mathrm{\mathbf{BRC}}$). The other seven memetic searches discovered new clusters (Table \ref{tab.7.clusters}). Six of them are small (maximum of 432 links) and not very well separated ($\Psi > 0.27$). Seed 196 (six sources) leads to a larger cluster, which has the same six top-300 sources as full members. They all deal with terrorism and form the top right corner of  $\mathrm{\mathbf{TR}}$ (violet in lower graph of Figure \ref{fig:graphs}, p.\;\pageref{fig:graphs}).

%%%%%%%%%%%%%%%%%%%%%%%%%%%%%%%%%
\subsection*{Search starting from seed 296} 
%%%%%%%%%%%%%%%%%%%%%%%%%%%%%%%%%
\label{sec:seed.296}
 \begin{figure}[!t]
\begin{center}
 \includegraphics[height=12cm]{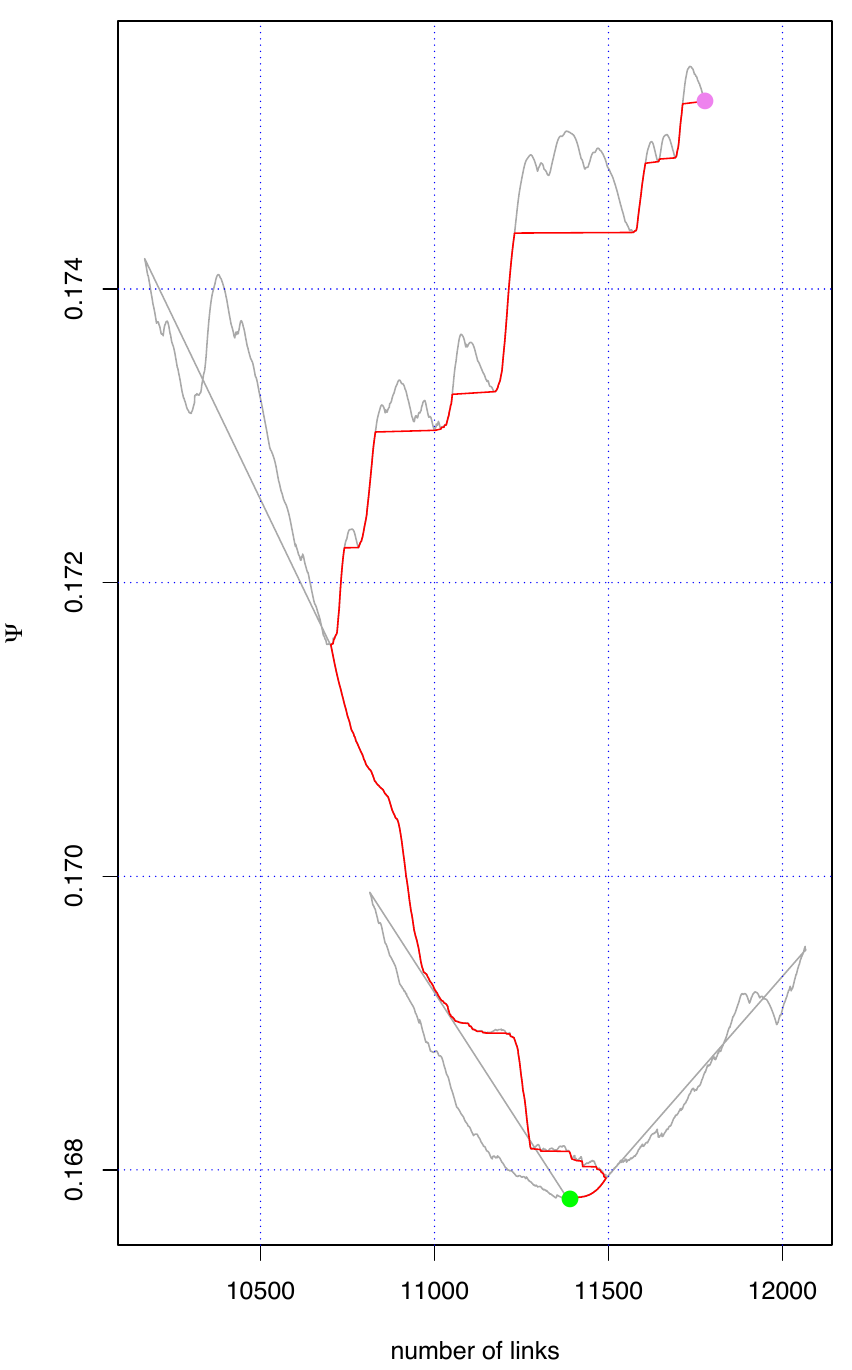}
\caption{Cost-size diagram of the local search starting from seed 296 and ending in $L_\mathrm{green}$ ($r=1/20$, cf.\;text)}
\label{fig:local.search}
\end{center}
\end{figure}

%%%%%%%%%%%%%%%%%%%%%%%
 \begin{figure}[h]
\begin{center}
 \includegraphics[height=13.7cm]{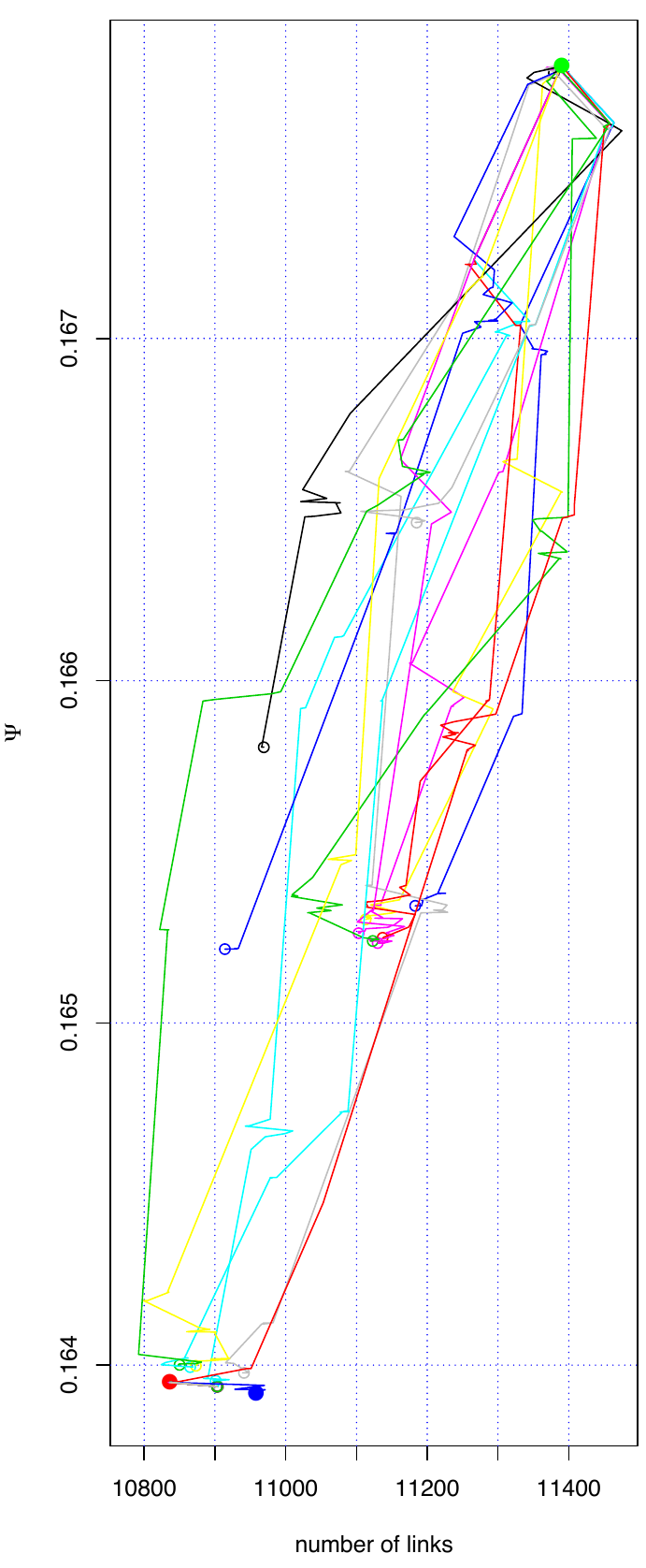}
 \caption{Cost-size diagram of the memetic search beginning at $L_\mathrm{green}$ and ending in cluster $\mathrm{\mathbf{R}}$ (blue point, $r=1/20$, cf.\;text)}
\label{fig:mem.search}
\end{center}
\end{figure}
%%%%%%%%%%%%%%%%%%%%%%%

To give an impression how PsiMinL finds lower places in the cost landscape, the height profile of 
the local search path that starts from seed 296 is displayed in Figure \ref{fig:local.search}. Greedy exclusion of links starts from the violet point and tunnels through several barriers (red and grey curves). Then the algorithm tries to exclude further links but the barrier  is to thick to tunnel through for resolution $r=1/20$ (grey curve on the left side). At 10,167 links,  exclusion ends and we turn back to the last cost minimum on the exclusion path (10,702 links, $\Psi\approx1.716$).   Now, greedy inclusion of links starts (red curve from left turning point  to right turning point). On the inclusion path there are only small barriers to tunnel through. Further inclusion is not successful and a second exclusion starts, which ends up in the green point after the algorithm failed to exclude more links ($|L_\mathrm{green}| = 11,390$, $\Psi(L_\mathrm{green}) \approx 0.1678$). 

In Figure \ref{fig:cost.size.27} (p.\;\pageref{fig:cost.size.27}), the upper part of the red line connecting seed 296 with cluster $\mathrm{\mathbf{R}}$ corresponds to the deterministic local search in Figure \ref{fig:local.search}.  
The lower part of the red line  corresponds to the memetic search visualised in Figure  \ref{fig:mem.search}. It started from $L_\mathrm{green}$ and ended up in $\mathrm{\mathbf{R}}$ (blue point in Figure\;\ref{fig:mem.search}) already for $r=1/20$. Increasing the tunnel length by increasing the resolution parameter up to $r=1/3$ did not give a better result. 

The different lines beginning at the green point correspond to 16 independent searches. Each line connects the current best clusters in a memetic evolution. The best result (red point) was reached by one search (red line). In a second phase, 16 evolutions started from the red point (using different eight best clusters as individuals for all 16 populations). One of them (blue line) reached  $\mathrm{\mathbf{R}}$ (blue point). In a third phase, PsiMinL tried to improve this cluster in further 16 searches but without success. Here the populations were initialised by mutating $\mathrm{\mathbf{R}}$ until eight different individuals were generated.

For seed 296, the memetic evolutions  including searches for all resolution levels needed 8.75\,h computer time with the R-package PsiMinL implemented under Ubuntu on a Dell machine with 56 CPUs (for further technical parameters, see next section).  The initial local search was made with an R-script (without using package PsiMinL) and needed 80\,s.

%%%%%%%%%%%%%%%%%%%%%%%%%%%%%%%
\subsection*{Computer running times\\and PsiMinL parameters}
%%%%%%%%%%%%%%%%%%%%%%%%%%%%%%%
\label{sec:run.times}

%%%%%%%%%%%%%%%%%%%%%%%%%%
 \begin{figure}[!b]
\begin{center}
 \includegraphics[height=6.7cm]{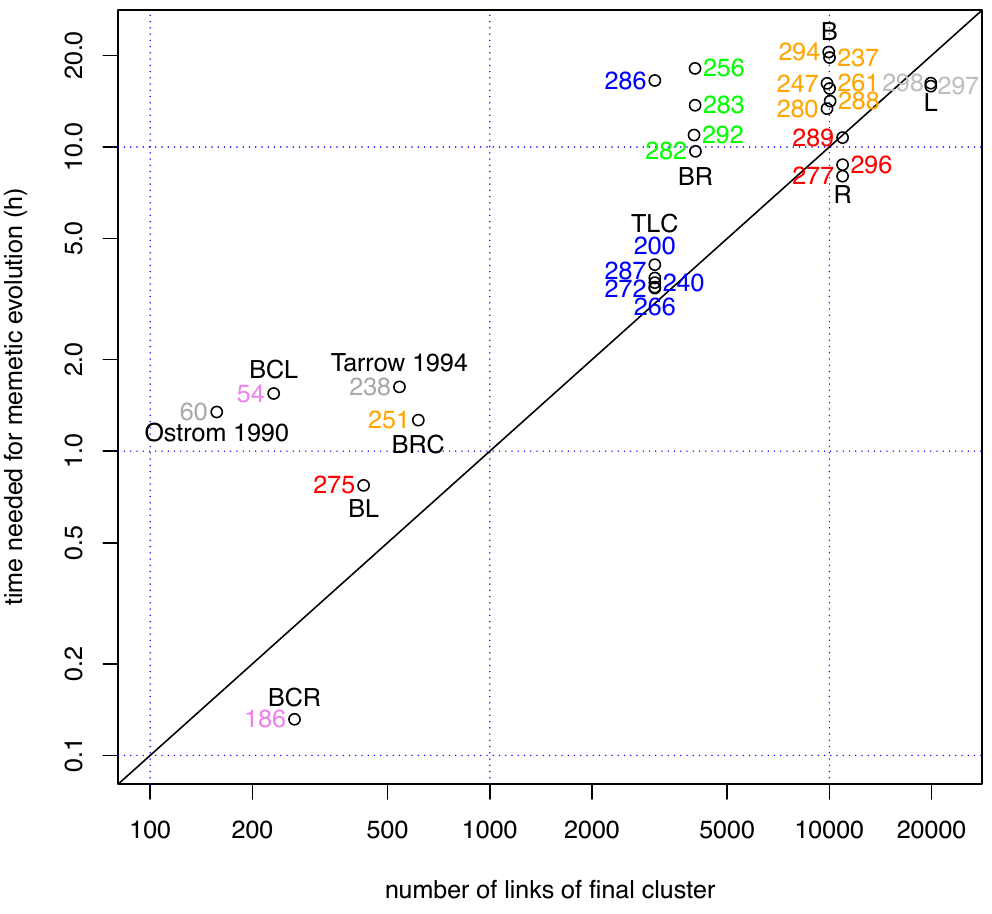}
 \caption{Computer running times  of  memetic searches (for the whole sequence of resolution parameters $r=1/20, \ldots, 1/3$) over size of the final cluster (numbers are seed numbers, their colours are the same as in Figure \ref{fig:cost.size.27}, p.\;\pageref{fig:cost.size.27}, cf.\;text)}
\label{fig:run.times}
\end{center}
\end{figure}
%%%%%%%%%%%%%%%%%%%%%%%%%

Figure \ref{fig:run.times} is a diagram of computer running times for different seeds. It shows that running times depend only partly on size of the final clusters.\footnote{Note, that some final clusters of the same colour differ slightly in size. As mentioned in section \ref{sec:exp.link} (p.\;\pageref{sec:exp.link}), in some cases, evolutions end up in slightly different variants of a cluster. The best one invalidates the other variants.} 

The straight line corresponds to a linear increase of time with size of one hour per 1000 links. 
For all but one of the small final clusters, PsiMinL needs more time per link. But for larger clusters, roughly speaking, the minimal time increases linearly with size and the maximal times are not extremely far away from minimal times (with the exception of seed 286, from which first the large cluster $\mathrm{\mathbf{TL}}$ is reached before evolution ends up in the smaller cluster $\mathrm{\mathbf{TLC}}$, cf.\;Figure\;\ref{fig:cost.size.27},\;p.\;\pageref{fig:cost.size.27}). How time scales with network size cannot be estimated without further experiments.
 
R-package PsiMinL was implemented on R-version 3.2.3 (2015-12-10, ``Wooden Christmas-Tree'')  under Ubuntu 16.04.2 LTS on a Dell machine with 56 CPUs (240\,GHz) and 126 GB RAM.
The values of technical PsiMinL parameters applied in the experiments are listed in Table\;\ref{tab:parameters} (p.\;\pageref{tab:parameters}).

Maximally, each PsiMinL process with eight individuals in a population needs eight CPUs. Thus, eight parallel processes running on 56 CPUs do not impede each other and,  for a network with about 30,000 links, the whole memetic search can be completed within a few days.

%%%%%%%%%%%%%%%%%%%%%%
\begin{figure}[!b]
\begin{center}
 \includegraphics[height=6.6cm]{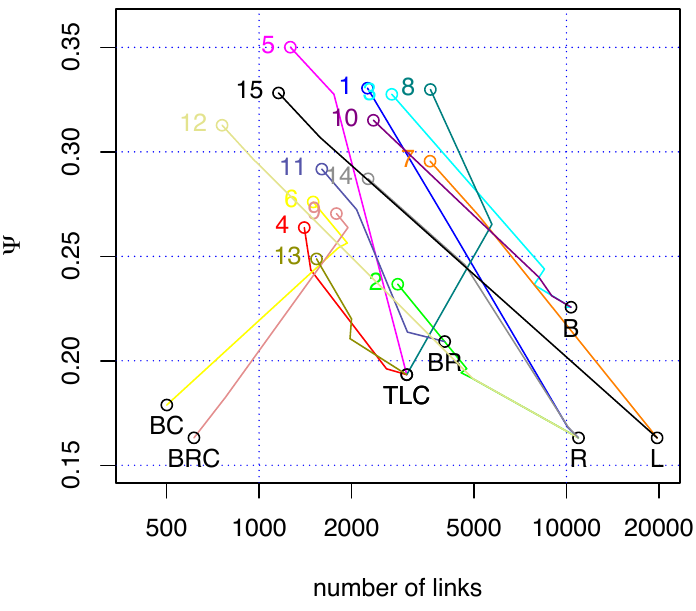}
 \caption{Cost-size diagram of 15 seed subgraphs and of steps towards seven clusters valid on all resolution levels $r\le1/3$}
\label{fig:cost.size}
\end{center}
\end{figure}

%%%%%%%%%%%%%%%
\begin{table}[!t]
\caption{Link clusters in incomplete network}
\begin{center}
\begin{tabular}{rlrr}
size&name of	& number&  		\\
rank& cluster & of links &  $\Psi$ \\
\hline
%\vspace{0.1cm}
1 & $\mathrm{\mathbf{L}}$ & 19,665 &  0.16315\\
2 & $\mathrm{\mathbf{R}}$ & 10,949 &  0.16315\\
3 &$\mathrm{\mathbf{B}}$ &10,350 & 0.22565\\
4  & $\mathrm{\mathbf{BR}}$ & 4,017 & 0.20924\\
 5   & $\mathrm{\mathbf{TLC}}$& 3,023 & 0.19339\\
  6  &$\mathrm{\mathbf{BRB}}$ & 891 &0.15864 \\
  7  & $\mathrm{\mathbf{BRC}}$ & 614 &0.16316\\
  8    & Cox 1981 & 532 & 0.27443 \\

 9   &  $\mathrm{\mathbf{BC}}$ & 501 & 0.17888 \\
 10       &Olson 1965& 302 &0.35075 \\

11&$\mathrm{\mathbf{BCR}}$& 266 & 0.14885 \\
12&North 1990 &263 &0.33023 \\

13&$\mathrm{\mathbf{BCL}}$& 231 & 0.21545 \\
\hline
 1 & $\mathrm{\mathbf{TL}}$ & 6,581 & 0.22266  \\
 2 &$\mathrm{\mathbf{TR}}$ &4,970 & 0.19142\\ 

\end{tabular}
\end{center}
\label{tab.clusters.old}
\end{table}%
%%%%%%%%%%%%%%%%%%%%%%%%%%%

%%%%%%%%%%%%%
\subsection*{Results with seeds \\from a disjoint partition}
\label{Ward-15}
%%%%%%%%%%%%%
In a first trial, I constructed seed subgraphs from a cut through the Ward dendrogram in Figure \ref{fig:dendro} (p.\;\pageref{fig:dendro}) that delivers 15 clusters.
On last resolution level $r=1/3$ the independent procedures starting from these seeds converged toward seven valid clusters, which are also valid for $r<1/3$ by definition (the five largest clusters, cluster $\mathrm{\mathbf{BRC}}$, and cluster $\mathrm{\mathbf{BC}}$ in Table \ref{tab.clusters.old}).

Figure \ref{fig:cost.size} shows  costs $\Psi$ and sizes  of all 15 Ward seed-subgraphs, of results of initial local searches and  memetic searches on intermediary resolution levels, and of their resulting clusters on final resolution level. Seed size does not correspond to cluster size and  different seeds end in the same cluster. 

\label{sec.comment}
The diagram visualises cost-size data of seeds and clusters obtained with an incomplete network with 30,614 citation links. After adding 221 links (mainly citations of Kant, 1795), I used all clusters as seeds for new memetic evolutions, which ended in very similar clusters. They are equal to some of the  clusters  in Table \ref{tab.clusters} (p.\;\pageref{tab.clusters}). The data of corresponding clusters in the incomplete network are shown in Table \ref{tab.clusters.old}. 

%%%%%%%%%%%%%%%%%%%%%%%
\subsection*{Comparing Ward with link clusters}
%%%%%%%%%%%%%%%%%%%%%%%%

I have compared any Ward cluster of top-300 sources with the set of sources $S(L)$ of any link cluster  $L$. Set $S(L)$  includes all cited sources that have more than half of their incoming citation links in link set $L$.

%%%%%%%%%%%%%%%
\begin{table}[!h]
\caption{All Ward clusters  $W$ best matching source sets $S(L)$ of link clusters $L$ ($o=|S(L)\cap W|=$ number of sources in overlap; $s=o/\sqrt{|S(L)||W|}=$ Salton index of $W$ and $S(L)$)}
\label{tab.match}
\begin{center}
\begin{tabular}{rrrrrr}
&&&&&\\
$W$ & $|W|$ & L & $|S(L)|$ &$o$&$s$\\
\hline
 & 1 & Olson  & 1 & 1 & 1.00\\
54 & 3 & $\mathrm{\mathbf{ BCL }}$ & 3 & 3 & 1.00\\
60 & 2 & Ostrom & 2 & 2 & 1.00\\
186 & 5 & $\mathrm{\mathbf{ BCR }}$ & 4 & 4 & 0.89\\
206 & 2 & North  & 2 & 2 & 1.00\\
231 & 6 &  Cox  & 5 & 4 & 0.73\\
238 & 6 & Tarrow  & 6 & 6 & 1.00\\
251 & 7 & $\mathrm{\mathbf{ BRC }}$ & 8 & 7 & 0.94\\
275 & 8 & $\mathrm{\mathbf{ BL }}$ & 5  & 5  & 0.79\\
276 & 21 & $\mathrm{\mathbf{ TLC }}$ & 29 & 19 & 0.77\\
286 & 35 & $\mathrm{\mathbf{ TL }}$ & 57 & 30 & 0.67\\
289 & 47 & $\mathrm{\mathbf{ TR }}$ & 46 & 39 & 0.84\\
292 & 50 & $\mathrm{\mathbf{ BR }}$ & 46 & 39 & 0.81\\
294 & 72 & $\mathrm{\mathbf{ B }}$ & 108 & 64 & 0.73\\
296 & 125 & $\mathrm{\mathbf{ R }}$ & 108 & 106 & 0.91\\
298 & 175 & $\mathrm{\mathbf{ L }}$ & 192 & 173 & 0.94\\
\label{tab.match}
\end{tabular}
\end{center}
\end{table}

%%%%%%%%%%%%

\label{sec:comp.Ward.Psi}
Going through the dendrogram in Figure \ref{fig:dendro} (p.\;\pageref{fig:dendro}) towards increasing numbers of branches, for each of all nontrivial Ward clusters $W$ I have searched for the link cluster $L$ with a source set $S(L)$ best matching $W$. $S(L)$ is best matching $W$ if their symmetric difference $(S(L)\cup W) - (S(L)\cap W)$ has minimum size. 

The link cluster obtained from Olson (1965) has only this paper in  $S(L)$, which is trivially matched with itself. 
Results are in Table \ref{tab.match}. All best matching seeds but seed 276 can be found in the dendrogram of Ward clustering in Figure\;\ref{fig:dendro}. Seed 276 and seed 272 are direct sub-clusters of seed 287, which can be found in the lower part of the dendrogram.

\newpage
\begin{table}[!p]
\begin{footnotesize}
\caption{List of 128 journals with numbers of papers with references (10- and 5-years periods)}
\label{tab.journals}

\begin{tabular}{>{\raggedright}p{3.7cm}rrr}
&&&\\
&&&\\
 & 2006& 2006 & 2011 \\
journal & --15& --10 & --15 \\
\hline
Afr. Secur. Rev. & 26 & 0 & 26 \\
All Azimuth & 8 & 0 & 8 \\
Alternatives & 197 & 99 & 98 \\
Am. Foreign Policy Interes. & 18 & 0 & 18 \\
Am. J. Int. Law & 741 & 392 & 349 \\
Am. J. Polit. Sci. & 632 & 304 & 328 \\
Am. Polit. Sci. Rev. & 461 & 224 & 237 \\
An. Mex. Derecho Int. & 151 & 59 & 92 \\
Asia Eur. J. & 264 & 134 & 130 \\
Asian J. WTO Int. Health Law Po. & 164 & 73 & 91 \\
Asian Perspec. & 210 & 78 & 132 \\
Aust. J. Int. Aff. & 538 & 261 & 277 \\
Biosecur. Bioterror. & 270 & 122 & 148 \\
Bol. Meridiano 47 & 29 & 0 & 29 \\
Br. J. Polit. Int. Relat. & 293 & 113 & 180 \\
Br. J. Polit. Sci. & 391 & 195 & 196 \\
Bull. Atom. Scient. & 423 & 163 & 260 \\
Camb. Rev. Int. Aff. & 460 & 156 & 304 \\
Canadian Foreign Policy & 20 & 0 & 20 \\
Chin. J. Int. Law & 251 & 86 & 165 \\
Chin. J. Int. Polit. & 135 & 55 & 80 \\
Columbia Int. & 175 & 77 & 98 \\
Columbia J. Transnatl. Law & 186 & 101 & 85 \\
Common Mark. Law Rev. & 1100 & 462 & 638 \\
Communist Post-Communist Stud. & 322 & 143 & 179 \\
Confl. Manage. Peace Sci. & 232 & 106 & 126 \\
Contemp. Southeast Asia & 132 & 0 & 132 \\
Contexto Int. & 182 & 75 & 107 \\
Coop. Confl. & 299 & 148 & 151 \\
Cornell Int. Law J. & 195 & 96 & 99 \\
Curr. Hist. & 87 & 46 & 41 \\
Emerg. Mark. Fin. Trade & 732 & 179 & 553 \\
Estudios Int. & 40 & 0 & 40 \\
Ethics Int. Aff. & 225 & 0 & 225 \\
Etud. Int. & 60 & 0 & 60 \\
Eur. J. Int. Law & 678 & 318 & 360 \\
Eur. J. Int. Relat. & 305 & 118 & 187 \\
Foreign Aff. & 2671 & 1258 & 1413 \\
Foreign Policy Anal. & 207 & 95 & 112 \\
Glob. Gov. & 372 & 161 & 211 \\
Glob. Policy & 371 & 51 & 320 \\
Glob. Soc. & 30 & 0 & 30 \\
Globalizations & 424 & 148 & 276 \\
India Q. & 42 & 0 & 42 \\
Int. Aff. & 3113 & 1491 & 1622 \\
Int. Interact. & 231 & 82 & 149 \\
Int. J. & 787 & 466 & 321 \\
Int. J. Confl. Violence & 155 & 63 & 92 \\
Int. J. Transitional Justice & 260 & 127 & 133 \\
Int. Organ. & 273 & 130 & 143 \\
Int. Peacekeeping & 364 & 110 & 254 \\
Int. Polit. Sociol. & 188 & 83 & 105 \\
Int. Polit. (Oslo) & 428 & 216 & 212 \\
Int. Politics & 313 & 116 & 197 \\
Int. Politik & 691 & 605 & 86 \\
Int. Relat. & 230 & 95 & 135 \\
Int. Relat. Asia-Pac. & 209 & 96 & 113 \\
Int. Secur. & 324 & 163 & 161 \\
Int. Stud. Perspect. & 239 & 85 & 154 \\
Int. Stud. Q. & 537 & 219 & 318 \\
Int. Stud. Rev. & 727 & 269 & 458 \\

\end{tabular}
\end{footnotesize}

\end{table}
\begin{table}
\begin{footnotesize}

\begin{tabular}{>{\raggedright}p{3.7cm}rrr}

&&&\\
Int. Theory & 141 & 42 & 99 \\
Issues Stud. & 173 & 132 & 41 \\
J. Cold War Stud. & 621 & 165 & 456 \\
J. Confl. Resolut. & 431 & 192 & 239 \\
J. Eur. Integr. & 207 & 0 & 207 \\
J. Hum. Rights & 237 & 81 & 156 \\
J. Int. Relat. Dev. & 209 & 98 & 111 \\
J. Int. Stud. -- JIS & 9 & 0 & 9 \\
J. Jpn. Int. Econ. & 303 & 144 & 159 \\
J. Marit. Law Commer. & 166 & 139 & 27 \\
J. Mil. Strateg. Stud. & 23 & 0 & 23 \\
J. Peace Res. & 1374 & 796 & 578 \\
J. Polit. & 807 & 403 & 404 \\
J. Strateg. Stud. & 492 & 276 & 216 \\
J. World Trade & 463 & 243 & 220 \\
J. Common Mark. Stud. & 1338 & 651 & 687 \\
Korea Obs. & 220 & 95 & 125 \\
Korean J. Def. Anal. & 304 & 144 & 160 \\
Latin Amer. Polit. Soc. & 573 & 256 & 317 \\
Mar. Pol. & 1640 & 560 & 1080 \\
Mediterr. Polit. & 271 & 112 & 159 \\
Mex. Cuenca Pac. & 15 & 0 & 15 \\
Middle East Policy & 574 & 280 & 294 \\
Migraciones Int. & 152 & 61 & 91 \\
Millennium -- J. Int. Stud. & 790 & 508 & 282 \\
Nav. War Coll. Rev. & 66 & 0 & 66 \\
New Polit. Econ. & 342 & 155 & 187 \\
North Korean Rev. & 123 & 69 & 54 \\
Ocean Dev. Int. Law & 188 & 98 & 90 \\
Pac. Focus & 173 & 78 & 95 \\
Pac. Rev. & 285 & 133 & 152 \\
Peacebuilding & 24 & 0 & 24 \\
Post-Sov. Aff. & 168 & 74 & 94 \\
Rel. Int. & 280 & 98 & 182 \\
Relac. Int. & 18 & 0 & 18 \\
Rev. Bras. Polit. Int. & 214 & 90 & 124 \\
Rev. CIDOB Afers Int. & 42 & 0 & 42 \\
Rev. Derecho Comun. Eur. & 203 & 166 & 37 \\
Rev. Int. Organ. & 185 & 69 & 116 \\
Rev. Int. Polit. Econ. & 385 & 183 & 202 \\
Rev. Int. Stud. & 535 & 224 & 311 \\
Rev. Rel. Int. Estrat. Sec. & 101 & 15 & 86 \\
Rev. UNISCI & 19 & 0 & 19 \\
Rev. World Econ. & 324 & 171 & 153 \\
S. Afr. J. Int. Aff. & 42 & 0 & 42 \\
Sci. Glob. Secur. & 10 & 0 & 10 \\
Secur. Chall. & 8 & 0 & 8 \\
Secur. Dialogue & 296 & 141 & 155 \\
Secur. Stud. & 241 & 109 & 132 \\
Si Somos Am. & 18 & 0 & 18 \\
Small War Insur. & 53 & 0 & 53 \\
Space Policy & 295 & 141 & 154 \\
Stability & 42 & 0 & 42 \\
Stanford J. Int. Law & 130 & 69 & 61 \\
Stud. Comp. Int. Dev. & 187 & 88 & 99 \\
Stud. Confl. Terror. & 547 & 285 & 262 \\
Survival & 1555 & 648 & 907 \\
Terror. Polit. Violence & 716 & 328 & 388 \\
Tla-Melaua & 19 & 0 & 19 \\
Uluslar. Iliskiler & 227 & 81 & 146 \\
Vestn. Mezhdunarodn. Org. & 22 & 0 & 22 \\
War Hist. & 688 & 327 & 361 \\
Wash. Q. & 403 & 199 & 204 \\
World Econ. & 854 & 417 & 437 \\
World Policy J. & 64 & 64 & 0 \\
World Polit. & 194 & 94 & 100 \\
World Trade Rev. & 289 & 105 & 184 \\
\hline
total & 56821 & 25896 & 30925 \\

\end{tabular}
\end{footnotesize}
\end{table}

\onecolumn
{
%%%%%%%%%%%%%%%%%%%
% table parameters
%%%%%%%%%%%%%%%%%%
\begin{table}[h]
\caption{Technical PsiMinL parameters, their values, and their meaning} 
\vspace{0.3cm}
\begin{tabularx}{\textwidth}{rrX}
&&\\
name	&			value &	meaning\\
\hline
&&\\

population size		& 8	&	number of individuals that produce offspring by crossovers and mutation\\
mutation variance	& 1	&	percentage of genes (links) that are randomly altered in mutation of the best individual\\
renewal variance	& 6	&	percentage of genes (links) that are randomly altered in a renewal mutation\\
mutation rate		& 4	&	number of mutants in each generation\\
number of crossovers & 4	&	number of gene combinations (unions and intersections) in each generation of the current best individual with other individuals\\
renewal period		&10	&	number of generations after that a renewal mutation is made if the best individual remains the same\\
stopping period	  & 100	&	number of generations after that evolution is stopped if the best individual remains the same\\
minimal period   &  200	&	minimal total number of generations of a population	\\

\end{tabularx}
\label{tab:parameters}
\end{table}

%%%%%%%%%%%%%%%%%%

\label{b.tab.1}
\begin{longtable}{ rr >{\raggedright}p{3.2cm}l p{7.9cm}}
\caption{Top 300 cited sources}\\
\label{e.tab.1}
rank & cit & (first) author & year & title and further bibliographic data\\
\hline
\hspace{0.1cm}
\endhead
\input{table-300-sources-v4}
\end{longtable}}

%%%%%%%%%%%%%%%%%%%%%%
\newpage
\twocolumn
\bibliography{networks}
\bibliographystyle{chicago}
\end{document}

%% file: table-300-sources-v4.tex
\textbf{200} & 139 & Abbott, Kenneth W. & 1998 & Why states act through formal international organizations. \textit{J Confl Resolut} \textbf{42}, 3 (with Snidal, D) \\
\textbf{229} & 125 & Abbott, Kenneth W. & 2000 & The concept of legalization. \textit{Int Organ} \textbf{54}, 401 \\
\textbf{112} & 188 & Abbott, Kenneth W. & 2000 & Hard and soft law in international governance. \textit{Int Organ} \textbf{54}, 421 (with Snidal, D) \\
278 & 93 & Acemoglu, Daron & 2001 & The colonial origins of comparative development: An empirical investigation. \textit{Am Econ Rev} \textbf{91}, 1369 (with Johnson, S; Robinson, JA) \\
152 & 164 & Acemoglu, Daron & 2006 & Economic Origins of Dictatorship and Democracy (with James A. Robinson) \\
\textbf{203} & 138 & Acharya, Amitav & 2001 & Constructing a security community in Southeast Asia: ASEAN and the problem of regional order \\
\textbf{120} & 180 & Acharya, Amitav & 2004 & How ideas spread: Whose norms matter? Norm localization and institutional change in Asian regionalism. \textit{Int Organ} \textbf{58}, 239 \\
\textbf{33} & 340 & Adler, Emanuel (ed.) & 1998 & Security Communities (with Michael Barnett) \\
\textbf{231} & 124 & Adler, Emanuel & 1997 & Seizing the Middle Ground: Constructivism in World Politics. \textit{European J Int Relat} \textbf{3}, 319 (with Peter M. Haas) \\
102 & 199 & Agamben, Giorgio & 1998 & Homo sacer: Sovereign power and bare life \\
222 & 128 & Agamben, Giorgio & 2003 & State of Exception \\
\textbf{270} & 99 & Alesina, Alberto & 2000 & Who Gives Foreign Aid to Whom and Why? \textit{J Econ Growth} \textbf{5}, 33 \\
282 & 89 & Alesina, Alberto & 2003 & Fractionalization. \textit{J Econ Growth} \textbf{8}, 155 (with Devleeschauwer, A; Easterly, W; Kurlat, S; Wacziarg, R) \\
\textbf{202} & 138 & Amsden, Alice H. & 1989 & Asia's Next Giant: South Korea and Late Industrialization \\
19 & 398 & Anderson, Benedict & 1983 & Imagined communities: Reflections on the origin and spread of nationalism \\
144 & 167 & Anderson, James E. & 2003 & Gravity with gravitas: A solution to the border puzzle. \textit{Am Econ Rev} \textbf{93}, 170 (with van Wincoop, E) \\
\textbf{266} & 102 & Appadurai, Arjun & 1996 & Modernity at Large: Cultural Dimensions of Globalization \\
166 & 156 & Arellano, Manuel & 1991 & Some tests of specification for panel data -- Monte-Carlo evidence and an application to employment equations. \textit{Rev Econ Stud} \textbf{58}, 277 (with Bond, S) \\
\textbf{87} & 222 & Axelrod, Robert M. & 1984 & The Evolution of Cooperation \\
\textbf{233} & 123 & Barnett, Michael N. & 1999 & The politics, power, and pathologies of international organizations. \textit{Int Organ} \textbf{53}, 699 (with Martha Finnemore) \\
\textbf{45} & 302 & Barnett, Michael & 2004 & Rules for the World: International Organizations in Global Politics (with Martha Finnemore) \\
\textbf{187} & 145 & Barnett, Michael & 2005 & Power in international politics. \textit{Int Organ} \textbf{59}, 39 (with Duvall, R) \\
272 & 98 & Baumgartner, Frank R. & 1993 & Agendas and Instability in American Politics (with Bryan D. Jones) \\
67 & 250 & Beck, Nathaniel & 1995 & What to do (and not to do) with time-series cross-section data. \textit{Am Polit Sci Rev} \textbf{89}, 634 (with Jonathan N. Katz) \\
15 & 423 & Beck, Nathaniel & 1998 & Taking time seriously: Time-series-cross-section analysis with a binary dependent variable. \textit{Am J Polit Sci} \textbf{42}, 1260 (with Katz, JN; Tucker, R) \\
\textbf{181} & 149 & Beck, Thorsten & 2001 & New tools in comparative political economy: The database of political institutions. \textit{World Bank Econ Rev} \textbf{15}, 165 (with Clarke, G; Groff, A; Keefer, P; Walsh, P) \\
167 & 155 & Beck, Ulrich & 1986 & Risikogesellschaft: auf dem Weg in eine andere Moderne (Risk Society: Towards a New Modernity) \\
298 & 65 & Benford, Robert D. & 2000 & Framing processes and social movements: An overview and assessment. \textit{Annu Rev Sociol} \textbf{26}, 611 (with Snow, DA) \\
\textbf{65} & 256 & Bennett, D. Scott & 2000 & EUGene: A conceptual manual. \textit{Int Interact} \textbf{26}, 179 (with Stam, AC) \\
\textbf{255} & 113 & Bigo, Didier & 2002 & Security and Immigration: Toward a Critique of the Governmentality of Unease. \textit{Alternatives} \textbf{27}, 63 \\
\textbf{295} & 69 & Blattman, Christopher & 2010 & Civil War. \textit{J Econ Lit} \textbf{48}, 3 (with Edward Miguel) \\
\textbf{185} & 146 & Bloom, Mia & 2005 & Dying to Kill: The Allure of Suicide Terror \\
219 & 128 & Blundell, Richard & 1998 & Initial conditions and moment restrictions in dynamic panel data models. \textit{J Econometrics} \textbf{87}, 115 (with Bond, S) \\
158 & 162 & Blyth, Mark & 2002 & Great Transformations. Economic Ideas and Institutional Change in the Twentieth Century \\
\textbf{286} & 82 & Boli, John (ed.) & 1999 &  Constructing world culture: international nongovernmental organizations since 1875 (with George M. Thomas) \\
\textbf{296} & 68 & Braithwaite, John & 2000 & Global Business Regulation \\
59 & 265 & Brambor, Thomas & 2006 & Understanding interaction models: Improving empirical analyses. \textit{Polit Anal} \textbf{14}, 63 (with Clark, WR; Golder, M) \\
\textbf{177} & 150 & Bremer, Stuart A. & 1992 & Dangerous dyads -- conditions affecting the likelihood of interstate war, 1816--1965. \textit{J Confl Resolut} \textbf{36}, 309 \\
\textbf{188} & 144 & Bull, Hedley (ed.) & 1984 & The Expansion of International Society \\
\textbf{9} & 565 & Bull, Hedley & 1977 & The anarchical society: a study of order in world politics \\
\textbf{130} & 172 & Buzan, Barry & 1991 & People, States and Fear: An Agenda for International Security Studies in the Post-Cold War Era \\
\textbf{12} & 454 & Buzan, Barry & 1998 & Security. A New Framework for Analysis (with Ole Waever and Jaap de Wilde) \\
\textbf{70} & 248 & Buzan, Barry & 2003 & Regions and Powers: The Structure of International Security (with Ole Waever) \\
\textbf{139} & 168 & Buzan, Barry & 2004 & From International to World Society?: English School Theory and Social Structure of Globalization \\
92 & 215 & Campbell, Angus & 1960 & The American Voter (with Philip E. Converse, Warren E. Miller, Donald E. Stokes) \\
\textbf{32} & 341 & Campbell, David & 1992 & Writing security: United States foreign policy and the politics of identity \\
170 & 155 & Carter, David B. & 2010 & Back to the Future: Modeling Time Dependence in Binary Data. \textit{Polit Anal} \textbf{18}, 271 (with Signorino, Curtis S.) \\
179 & 149 & Castells, Manuel & 1996 & The rise of the network society. The information age: economy, society and culture \\
\textbf{267} & 101 & Chayes, Abram & 1993 & On Compliance. \textit{Int Organ} \textbf{47}, 175 \\
\textbf{153} & 163 & Chayes, Abram & 1995 & The New Sovereignty: Compliance with International Regulatory Agreements (with Antonia Handler Chayes) \\
\textbf{194} & 141 & Checkel, Jeffrey T. & 1998 & The Constructivist Turn in International Relations Theory. \textit{World Polit} \textbf{50}, 324 \\
\textbf{162} & 158 & Checkel, Jeffrey T. & 2001 & Why comply? Social learning and European identity change. \textit{Int Organ} \textbf{55}, 553 \\
\textbf{218} & 129 & Checkel, Jeffrey T. & 2005 & International institutions and socialization in Europe: Introduction and framework. \textit{Int Organ} \textbf{59}, 801 \\
226 & 127 & Cheibub, Jose Antonio & 2010 & Democracy and dictatorship revisited. \textit{Public Choice} \textbf{143}, 67 (with Gandhi, Jennifer; Vreeland, James Raymond) \\
\textbf{262} & 107 & Collier, Paul & 1998 & On economic causes of civil war. \textit{Oxf Econ Pap -- New Ser} \textbf{50}, 563 (with Hoeffler, A) \\
\textbf{131} & 172 & Collier, Paul & 2003 & Breaking the conflict trap: civil war and development policy (with V.L. Elliott, Havard Hegre, Anke Hoeffler, Marta Reynal-Querol, Nicholas Sambanis) \\
\textbf{24} & 374 & Collier, Paul & 2004 & Greed and grievance in civil war. \textit{Oxf Econ Pap -- New Ser} \textbf{56}, 563 (with Hoeffler, A) \\
\textbf{210} & 134 & Collier, Paul & 2007 & The Bottom Billion: Why the Poorest Countries are Failing and What Can Be Done About It \\
128 & 172 & Converse, Philip E. & 1964 & The Nature of Belief Systems in Mass Publics. In: David Apter (ed.): Ideology and Discontent, P206 \\
\textbf{90} & 219 & Cox, Robert W. & 1981 & Social Forces, States and World Orders: Beyond International Relations Theory. \textit{Millenium J Int Stud} \textbf{10}, 126 \\
100 & 201 & Dahl, Robert A. & 1971 & Polyarchy: participation and opposition \\
\textbf{134} & 169 & de Mesquita, Bruce Bueno & 1999 & An institutional explanation of the democratic peace. \textit{Am Polit Sci Rev} \textbf{93}, 791 (with Morrow, JD; Siverson, RM; Smith, A) \\
10 & 489 & de Mesquita, Bruce Bueno & 2003 & The Logic of Political Survival \\
205 & 137 & Dean, Mitchell & 1999 & Governmentality: Power and Rule in Modern Society \\
\textbf{56} & 275 & Deutsch, Karl W. & 1957 & Political Community and the North Atlantic Area: International Organization in the Light of Historical Experience \\
236 & 122 & DiMaggio, Paul J. & 1983 & The iron cage revisited -- institutional isomorphism and collective rationality in organizational fields. \textit{Am Sociol Rev} \textbf{48}, 147 (with Powell, WW) \\
27 & 359 & Downs, Anthony & 1957 & An Economic Theory of Democracy \\
\textbf{115} & 185 & Downs, George W. & 1996 & Is the good news about compliance good news about cooperation? \textit{Int Organ} \textbf{50}, 379 \\
\textbf{183} & 148 & Doyle, Michael W. & 1983 & Kant, liberal legacies, and foreign affairs. \textit{Philos Public Aff} \textbf{12}, 205\\
\textbf{172} & 153 & Doyle, Michael W. & 1986 & Liberalism and world-politics. \textit{Am Polit Sci Rev} \textbf{80}, 1151 \\
\textbf{214} & 131 & Doyle, Michael W. & 2000 & International peacebuilding: A theoretical and quantitative analysis. \textit{Am Polit Sci Rev} \textbf{94}, 779 (with Sambanis, N) \\
\textbf{164} & 158 & Doyle, Michael W. & 2006 & Making War and Building Peace: United Nations Peace Operations (with Nicholas Sambanis) \\
\textbf{216} & 130 & Drezner, Daniel W. & 2007 & All Politics Is Global: Explaining International Regulatory Regimes \\
\textbf{81} & 226 & Duffield, Mark & 2001 & Global Governance and the New Wars: The Merging of Development and Security \\
\textbf{189} & 144 & Duffield, Mark & 2007 & Development, Security and Unending War: Governing the World of Peoples \\
281 & 90 & Escobar, Arturo & 1995 & Encountering Development: The Making and Unmaking of the Third World \\
191 & 141 & Esping-Andersen, Gosta & 1990 & The Three Worlds of Welfare Capitalism \\
168 & 155 & Evans, Peter B. & 1995 & Embedded Autonomy: State and Industrial Transformation \\
\textbf{41} & 316 & Fearon, James D. & 1994 & Domestic political audiences and the escalation of international disputes. \textit{Am Polit Sci Rev} \textbf{88}, 577 \\
\textbf{14} & 438 & Fearon, James D. & 1995 & Rationalist explanations for war. \textit{Int Organ} \textbf{49}, 379 \\
\textbf{193} & 141 & Fearon, James D. & 1997 & Signaling foreign policy interests -- Tying hands versus sinking costs. \textit{J Confl Resolut} \textbf{47}, 68 \\
\textbf{207} & 135 & Fearon, James D. & 1998 & Bargaining, enforcement, and international cooperation. \textit{Int Organ} \textbf{52}, 269 \\
\textbf{6} & 692 & Fearon, James D. & 2003 & Ethnicity, insurgency, and civil war. \textit{Am Polit Sci Rev} \textbf{97}, 75 \\
\textbf{146} & 166 & Fearon, James D. & 2004 & Why do some civil wars last so much longer than others? \textit{J Peace Res} \textbf{41}, 275 \\
\textbf{77} & 236 & Finnemore, Martha & 1996 & National interests in international society \\
\textbf{8} & 580 & Finnemore, Martha & 1998 & International norm dynamics and political change. \textit{Int Organ} \textbf{52}, 887 (with Sikkink, K) \\
\textbf{148} & 165 & Finnemore, Martha & 2003 & The Purpose of Intervention: Changing Beliefs About the Use of Force \\
196 & 140 & Fiorina, Morris P. & 1981 & Retrospective Voting in American National Elections \\
68 & 248 & Foucault, Michel & 1975 & Surveiller et punir: Naissance de la prison (Discipline and punish: the birth of the prison) \\
138 & 168 & Foucault, Michel & 1980 & Power/Knowledge: Selected Interviews and Other Writings 1972--1977 (ed. by Colin Gordon) \\
192 & 141 & Foucault, Michel & 1991 & The Foucault Effect: Studies in Governmentality: With Two Lectures by and an Interview with Michel Foucault, ed. by Graham Burchell, Colin Gordon, and Peter Miller \\
119 & 181 & Foucault, Michel & 1997 & Il faut défendre la société: cours au Collège de France (1975--1976). (Society must be defended: lectures at the Collège de France, 1975--76) \\
132 & 172 & Foucault, Michel & 2004 & Sécurité, Territoire, Population: Cours au Collège de France 1977--1978 (Security, Territory, Population: Lectures at the College de France, 1977--1978) \\
\textbf{171} & 154 & Franck, Thomas M. & 1990 & The Power of Legitimacy Among Nations \\
\textbf{74} & 244 & Fukuyama, Francis & 1992 & The End of History and the Last Man \\
264 & 103 & Geertz, Clifford & 1973 & The Interpretation of Cultures: Selected Essays \\
129 & 172 & Gellner, Ernest & 1983 & Nations and nationalism \\
26 & 361 & George, Alexander L. & 2005 & Case Studies and Theory Development in the Social Sciences (with Andrew Bennett) \\
\textbf{99} & 205 & Ghosn, Faten & 2004 & The MID3 data set, 1993--2001: Procedures, coding rules, and description. \textit{Confl Manage Peace Sci} \textbf{21}, 133 (with Palmer, G; Bremer, SA) \\
\textbf{224} & 127 & Gibler, Douglas M. & 2004 & Measuring Alliances: the Correlates of War Formal Interstate Alliance Dataset, 1816–-2000. \textit{J Peace Res} \textbf{41}, 211 \\
206 & 135 & Giddens, Anthony & 1984 & The constitution of society: outline of the theory of structuration \\
\textbf{11} & 470 & Gilpin, Robert & 1981 & War and change in world politics \\
\textbf{178} & 149 & Gilpin, Robert & 1987 & The political economy of international relations \\
\textbf{49} & 291 & Gleditsch, Kristian Skrede & 2002 & Expanded trade and GDP data. \textit{J Confl Resolut} \textbf{46}, 712 \\
\textbf{18} & 399 & Gleditsch, Nils Petter & 2002 & Armed conflict 1946--2001: A new dataset. \textit{J Peace Res} \textbf{39}, 615 (with Wallensteen, P; Eriksson, M; Sollenberg, M; Strand, H) \\
\textbf{201} & 139 & Goldsmith, Jack L. & 2005 & The Limits of International Law \\
\textbf{290} & 78 & Goldstein, Judith (ed.) & 1993 & Ideas and Foreign Policy: Beliefs, Institutions, and Political Change (with Robert O. Keohane) \\
\textbf{252} & 115 & Gourevitch, Peter & 1978 & The Second Image Reversed: The International Sources of Domestic Politics. \textit{Int Organ} \textbf{32}, 881 \\
127 & 175 & Gramsci, Antonio & 1971 & Selections from the Prison Notebooks \\
276 & 93 & Granovetter, Mark S. & 1973 & The Strength of Weak Ties. \textit{Am J Sociol} \textbf{78}, 1360 \\
\textbf{245} & 118 & Grant, Ruth W. & 2005 & Accountability and abuses of power in world politics. \textit{Am Polit Sci Rev} \textbf{99}, 29 (with Keohane, Robert O.) \\
\textbf{263} & 105 & Grieco, Joseph M. & 1988 & Anarchy and the limits of cooperation: a realist critique of the newest liberal institutionalism. \textit{Int Organ} \textbf{42}, 485 \\
\textbf{66} & 252 & Gurr, Ted Robert & 1970 & Why men rebel \\
\textbf{98} & 205 & Haas, Ernst B. & 1958 & The uniting of Europe: political, social and economic forces, 1950--1957 \\
\textbf{62} & 261 & Haas, Peter M. & 1992 & Epistemic communities and international-policy coordination -- introduction. \textit{Int Organ} \textbf{46}, 1 \\
108 & 192 & Habermas, Jürgen & 1992 & Faktizität und Geltung -- Beiträge zur Diskurstheorie des Rechts und des demokratischen Rechtsstaats (Between Facts and Norms, Contributions to a Discourse Theory of Law and Democracy) \\
\textbf{284} & 87 & Hafner-Burton, Emilie M. & 2005 & Human rights in a globalizing world: The paradox of empty promises. \textit{Am J Sociol} \textbf{110}, 1373 \\
259 & 109 & Hall, Peter A. & 1993 & Policy paradigms, social-learning, and the state -- the case of economic policy-making in britain. \textit{Comp Polit} \textbf{25}, 275 \\
265 & 102 & Hall, Peter A. & 1996 & Political science and the three new institutionalisms. \textit{Polit Stud} \textbf{44}, 936 (with Taylor, RCR) \\
64 & 259 & Hall, Peter A. & 2001 & Varieties of capitalism: The institutional foundations of comparative advantage (with David Soskice) \\
\textbf{145} & 167 & Hansen, Lene & 2006 & Security as practice: discourse analysis and the Bosnian war \\
235 & 122 & Hardin, Garrett & 1968 & The Tragedy of the Commons. \textit{Science} \textbf{162}, 1244 \\
\textbf{113} & 188 & Hardt, Michael & 2000 & Empire. Globalization as a new Roman order, awaiting its early Christians (with Antonio Negri) \\
\textbf{197} & 140 & Harvey, David & 2003 & The New Imperialism \\
157 & 163 & Harvey, David & 2005 & A Brief History of Neoliberalism \\
\textbf{234} & 123 & Hathaway, Oona A. & 2002 & Do human rights treaties make a difference? \textit{Yale Law J} \textbf{111}, 1935 \\
\textbf{88} & 221 & Hawkins, Darren G. (ed.) & 2006 & Delegation and Agency in International Organizations (with David A. Lake, Daniel L. Nielson, Michael J. Tierney) \\
\textbf{174} & 153 & Hegre, Havard & 2006 & Sensitivity analysis of empirical results on civil war onset. \textit{J Confl Resolut} \textbf{50}, 508 (with Sambanis, Nicholas) \\
\textbf{95} & 210 & Hegre, Havard & 2001 & Toward a democratic civil peace? Democracy, political change, and civil war, 1816--1992. \textit{Am Polit Sci Rev} \textbf{95}, 33 \\
\textbf{176} & 151 & Held, David & 1995 & Democracy and the global order \\
\textbf{150} & 164 & Held, David & 1999 & Global Transformations: Politics, Economics and Culture \\
\textbf{48} & 291 & Hoffman, Bruce & 1998 & Inside terrorism \\
\textbf{287} & 80 & Homer-Dixon, Thomas F. & 1999 & Environment, Scarcity, and Violence \\
\textbf{297} & 68 & Hooghe, Liesbet & 2009 & A Postfunctionalist Theory of European Integration: From Permissive Consensus to Constraining Dissensus. \textit{Br J Polit Sci} \textbf{39}, 1 (with Marks, Gary) \\
\textbf{251} & 116 & Hopf, Ted & 2002 & Social Construction of International Politics: Identities and Foreign Policies, Moscow, 1955 and 1999 \\
\textbf{58} & 268 & Horowitz, Donald L. & 1985 & Ethnic Groups in Conflict \\
61 & 262 & Huntington, Samuel P. & 1968 & Political order in changing societies \\
36 & 330 & Huntington, Samuel P. & 1991 & The Third Wave: Democratization in the Late Twentieth Century \\
\textbf{22} & 390 & Huntington, Samuel P. & 1996 & The clash of civilizations and the remaking of world order \\
\textbf{228} & 126 & Hurd, Ian & 1999 & Legitimacy and authority in international politics. \textit{Int Organ} \textbf{53}, 379 \\
\textbf{225} & 127 & Huysmans, Jef & 2006 & The Politics of Insecurity: Fear, Migration and Asylum in the EU \\
\textbf{37} & 330 & Ikenberry, G. John & 2001 & After Victory. Institutions, Strategic Restraint, and the Rebuilding of Order After Major Wars \\
\textbf{126} & 176 & Jaggers, Keith & 1995 & Tracking democracy 3rd-wave with the polity-iii data. \textit{J Peace Res} \textbf{32}, 469 (with Gurr, TR) \\
\textbf{28} & 359 & Jervis, Robert & 1976 & Perception and Misperception in International Politics \\
\textbf{121} & 179 & Jervis, Robert & 1978 & Cooperation under security dilemma. \textit{World Polit} \textbf{30}, 167 \\
\textbf{208} & 135 & Johnston, Alastair Iain & 2001 & Treating, international institutions as social environments. \textit{Int Stud Q} \textbf{45}, 487 \\
\textbf{118} & 181 & Jones, Daniel M. & 1996 & Militarized interstate disputes, 1816--1992: Rationale, coding rules, and empirical patterns. \textit{Confl Manage Peace Sci} \textbf{15}, 163 \\
\textbf{151} & 164 & Kagan, Robert & 2003 & Of Paradise and Power: America and Europe in the New World Order \\
257 & 112 & Kahneman, Daniel & 1979 & Prospect theory -- analysis of decision under risk. \textit{Econometrica} \textbf{47}, 263 (with Tversky, A) \\
\textbf{40} & 317 & Kaldor, Mary & 1999 &  New and Old Wars. Organized Violence in a Global Era \\
\textbf{39} & 321 & Kalyvas, Stathis N. & 2006 & The Logic of Violence in Civil War \\
\textbf{136} & 168 & Kant, Immanuel & 1795 & Zum ewigen Frieden. Ein philosophischer Entwurf (Perpetual Peace: A Philosophical Sketch) \\
124 & 177 & Kant, Immanuel & 1970 &  Kant: Political Writings, ed.\;by H.S.\;Reiss, transl. by H.B.\;Nisbet \\
\textbf{31} & 345 & Katzenstein, Peter J. (ed.) & 1996 & The Culture of National Security: Norms and Identity in World Politics \\
\textbf{7} & 674 & Keck, Margaret E. & 1998 & Activists beyond Borders. Advocacy Networks in International Politics (with Sikkink, Kathryn) \\
\textbf{72} & 244 & Kennedy, Paul & 1987 & The Rise and Fall of the Great Powers: Economic Change and Military Conflict from 1500 to 2000 \\
\textbf{17} & 409 & Keohane, Robert O. & 1977 & Power and interdependence: world politics in transition (with Joseph S. Nye, Jr.) \\
\textbf{3} & 724 & Keohane, Robert O. & 1984 & After Hegemony: Cooperation and Discord in the World Political Economy \\
\textbf{260} & 109 & Keohane, Robert O. & 1995 & The Promise of Institutionalist Theory. \textit{Int Secur} \textbf{20}, 39 \\
\textbf{137} & 168 & Kindleberger, Charles P. & 1973 & The World in Depression, 1929--1939 \\
34 & 339 & King, Gary & 1994 & Designing Social Enquiry: Scientific Inference in Qualitative Research (with  Robert Keohane and Sidney Verba) \\
35 & 337 & King, Gary & 2000 & Making the most of statistical analyses: Improving interpretation and presentation. \textit{Am J Polit Sci} \textbf{44}, 347 (with Tomz, M; Wittenberg, J) \\
114 & 186 & Kingdon, John W. & 1984 & Agendas, alternatives, and public policies \\
\textbf{275} & 94 & Koh, Harold Hongju & 1997 & Why Do Nations Obey International Law? \textit{Yale Law J} \textbf{106}, 2599 \\
\textbf{79} & 235 & Koremenos, Barbara & 2001 & The rational design of international institutions. \textit{Int Organ} \textbf{55}, 761 (with Lipson, C; Snidal, D) \\
\textbf{50} & 289 & Krasner, Stephen D. (ed.) & 1983 & International Regimes \\
\textbf{54} & 277 & Krasner, Stephen D. & 1999 & Sovereignty: Organized Hypocrisy \\
\textbf{140} & 167 & Kratochwil, Friedrich V. & 1989 & Rules, Norms, and Decisions: On the Conditions of Practical and Legal Reasoning in International Relations and Domestic Affairs \\
293 & 73 & La Porta, Rafael & 1999 & The quality of government. \textit{J Law Econ Organ} \textbf{15}, 222 \\
\textbf{246} & 118 & Lacina, Bethany & 2005 & Monitoring trends in global combat: A new dataset of battle deaths. \textit{Eur J Popul -- Rev Eur Demogr} \textbf{21}, 145 (with Gleditsch, NP) \\
\textbf{256} & 113 & Li, Quan & 2005 & Does Democracy Promote or Reduce Transnational Terrorist Incidents? \textit{J Confl Resolut} \textbf{49}, 278 \\
268 & 100 & Lijphart, Arend & 1977 & Democracy in Plural Societies. A comparative Exploration \\
122 & 179 & Lijphart, Arend & 1999 & Patterns of Democracy. Government Forms and Performance in Thirty-Six Countries \\
107 & 193 & Linz, Juan J. & 1996 & Problems of Democratic Transition and Consolidation: Southern Europe, South America, and Post-Communist Europe (with Alfred Stepan) \\
182 & 148 & Lipset, Seymour Martin & 1959 & Some social requisites of democracy -- economic-development and political legitimacy. \textit{Am Polit Sci Rev} \textbf{53}, 69 \\
\textbf{44} & 302 & Manners, Ian & 2002 & Normative power Europe: A contradiction in terms? \textit{J Common Mark Stud} \textbf{40}, 235 \\
\textbf{247} & 118 & Mansfield, Edward D. & 2005 & Electing To Fight: Why Emerging Democracies Go To War (BCSIA Studies in International Security) (with Jack L. Snyder) \\
\textbf{97} & 208 & Maoz, Zeev & 1993 & Normative and structural causes of democratic peace, 1946--1986. \textit{Am Polit Sci Rev} \textbf{87}, 624 (with Russett, B) \\
133 & 171 & March, James G. & 1989 & Rediscovering Institutions: The Organizational Basis of Politics (with Johan P. Olsen) \\
\textbf{155} & 163 & March, James G. & 1998 & The institutional dynamics of international political orders. \textit{Int Organ} \textbf{52}, 943 (with Olsen, JP) \\
243 & 118 & Marx, Karl & 1867 & Das Kapital. Kritik der politischen Ökonomie (Capital. A Critique of Political Economy) \\
\textbf{269} & 99 & Mattli, Walter & 1999 & The Logic of Regional Integration. Europe and Beyond \\
160 & 159 & Mayhew, David R. & 1974 & Congress, The Electoral Connection \\
285 & 84 & McAdam, Doug (ed.) & 1996 & Comparative Perspectives on Social Movements: Political Opportunities, Mobilizing Structures, and Cultural Framings (with John D. McCarthy, Mayer N. Zald) \\
204 & 138 & McAdam, Doug & 2001 & Dynamics of Contention (with Sidney Tarrow, Charles Tilly) \\
292 & 74 & McCarthy, John D. & 1977 & Resource mobilization and social-movements -- partial theory. \textit{Am J Sociol} \textbf{82}, 1212 (with Zald, MN) \\
\textbf{109} & 191 & Mearsheimer, John J. & 1990 & Back to the Future: Instability in Europe after the Cold War. \textit{Int Secur} \textbf{15}, 5 \\
\textbf{63} & 261 & Mearsheimer, John J. & 1994 & The False Promise of International Institutions. \textit{Int Secur} \textbf{19}, 5 \\
\textbf{5} & 699 & Mearsheimer, John J. & 2001 & The Tragedy of Great Power Politics \\
89 & 220 & Melitz, Marc J. & 2003 & The impact of trade on intra-industry reallocations and aggregate industry productivity. \textit{Econometrica} \textbf{71}, 1695 \\
289 & 79 & Meyer, John W. & 1977 & Institutionalized organizations -- formal-structure as myth and ceremony. \textit{Am J Sociol} \textbf{83}, 340 (with Rowan, B) \\
\textbf{212} & 132 & Meyer, John W. & 1997 & World society and the nation-state. \textit{Am J Sociol} \textbf{103}, 144 (with Boli, J; Thomas, GM; Ramirez, FO) \\
\textbf{291} & 77 & Miguel, Edward & 2004 & Economic shocks and civil conflict: An instrumental variables approach. \textit{J Polit Econ} \textbf{112}, 725 (with Satyanath, S; Sergenti, E) \\
\textbf{147} & 165 & Milner, Helen V. & 1997 & Interests, institutions and information. Domestic politics and international relations \\
\textbf{283} & 88 & Moravcsik, Andrew & 1993 & Preferences and Power in the European Community: A Liberal Intergovernmentalist Approach. \textit{J Common Mark Stud} \textbf{31}, 473 \\
\textbf{104} & 198 & Moravcsik, Andrew & 1997 & Taking preferences seriously: A liberal theory of international politics. \textit{Int Organ} \textbf{51}, 513 \\
\textbf{53} & 284 & Moravcsik, Andrew & 1998 & The Choice for Europe: Social Purpose and State Power from Messina to Maastricht \\
\textbf{209} & 134 & Moravcsik, Andrew & 2000 & The origins of human rights regimes: Democratic delegation in postwar Europe. \textit{Int Organ} \textbf{54}, 217 \\
\textbf{4} & 712 & Morgenthau, Hans J. & 1948 & Politics among Nations: The Struggle for Power and Peace \\
\textbf{300} & 42 & Mueller, John E. & 1973 & War, Presidents, and Public Opinion \\
248 & 117 & North, Douglass C. & 1989 & Constitutions and Commitment -- The Evolution of Institutions Governing Public Choice In 17th-century England. \textit{J Econ Hist} \textbf{49}, 803 \\
38 & 321 & North, Douglass Cecil & 1990 & Institutions, Institutional Change and Economic Performance \\
\textbf{249} & 117 & Nye, Jr Joseph S. & 1990 & Bound to Lead: The Changing Nature of American Power \\
\textbf{30} & 352 & Nye, Jr Joseph S. & 2004 & Soft Power: The Means to Sucess in World Politics \\
101 & 201 & O'Donnell, Guillermo & 1986 & Transitions from Authoritarian Rule: Tentative Conclusions about Uncertain Democracies (with Philippe C. Schmitter, Laurence Whitehead) \\
16 & 422 & Olson, Mancur & 1965 & The Logic of Collective Action: Public Goods and the Theory of Groups \\
241 & 119 & Olson, Mancur & 1993 & Dictatorship, Democracy, and Development. \textit{Am Polit Sci Rev} \textbf{87}, 567 \\
\textbf{180} & 149 & Oneal, John R. & 1997 & The classical liberals were right: Democracy, interdependence, and conflict, 1950--1985. \textit{Int Stud Q} \textbf{41}, 267 (with Russett, BM) \\
\textbf{141} & 167 & Onuf, Nicholas Greenwood & 1989 & World of Our Making: Rules and Rule in Social Theory and International Relations \\
\textbf{125} & 176 & Organski, Abramo F. K. & 1980 & The War Ledger (with Jacek Kugler) \\
103 & 198 & Ostrom, Elinor & 1990 & Governing the commons: the evolution of institutions for collective action \\
240 & 120 & Page, Benjamin I. & 1992 & The Rational Public: Fifty Years of Trends in Americans’ Policy Preferences (with Robert Y. Shapiro) \\
\textbf{173} & 153 & Pape, Robert A. & 2003 & The strategic logic of suicide terrorism. \textit{Am Polit Sci Rev} \textbf{97}, 343 \\
\textbf{94} & 211 & Pape, Robert A. & 2005 & Dying to win: The strategic logic of suicide terrorism \\
\textbf{163} & 158 & Pape, Robert A. & 2005 & Soft balancing against the United States. \textit{Int Secur} \textbf{30}, 7 \\
\textbf{75} & 242 & Paris, Roland & 2004 & At War's End: Building Peace after Civil Conflict \\
250 & 116 & Pierson, Paul & 2000 & Increasing returns, path dependence, and the study of politics. \textit{Am Polit Sci Rev} \textbf{94}, 251 \\
156 & 163 & Pierson, Paul & 2004 & Politics in time: history, institutions, and social analysis \\
\textbf{199} & 139 & Poe, Steven C. & 1994 & Repression of human-rights to personal integrity in the 1980s -- a global analysis. \textit{Am Polit Sci Rev} \textbf{88}, 853 (with Tate, CN) \\
\textbf{220} & 128 & Poe, Steven C. & 1999 & Repression of the human right to personal integrity revisited: A global cross-national study covering the years 1976--1993. \textit{Int Stud Q} \textbf{43}, 291 \\
80 & 234 & Polanyi, Karl & 1944 & The Great Transformation \\
\textbf{244} & 118 & Posen, Barry R. & 1993 & The security dilemma and ethnic conflict. \textit{Survival} \textbf{35}, 27 \\
\textbf{242} & 119 & Powell, Robert & 2006 & War as a commitment problem. \textit{Int Organ} \textbf{60}, 169 \\
165 & 157 & Przeworski, Adam & 1991 & Democracy and the Market: Political and Economic Reforms in Eastern Europe and Latin America \\
43 & 304 & Przeworski, Adam & 2000 & Democracy and development: political institutions and well-being in the world, 1950--1990 (with M.E. Alvarez, J.A. Cheibub, F. Limongi) \\
\textbf{13} & 438 & Putnam, Robert D. & 1988 & Diplomacy and Domestic Politics -- The Logic of 2-level Games. \textit{Int Organ} \textbf{42}, 427 \\
82 & 224 & Putnam, Robert D. & 1993 & Making democracy work: civic traditions in modern Italy \\
116 & 184 & Putnam, Robert D. & 2000 & Bowling Alone: The Collapse and Revival of American Community \\
\textbf{288} & 80 & Raustiala, Kal & 2004 & The Regime Complex for Plant Genetic Resources. \textit{Int Organ} \textbf{58}, 277 (with David G. Victor) \\
76 & 237 & Rawls, John & 1971 & A Theory of Justice \\
123 & 179 & Rawls, John & 1999 & The Law of Peoples. With ``The Idea of Public Reason Revisited'' \\
\textbf{135} & 169 & Reiter, Dan & 2002 & Democracies at War (with Allan C. Stam) \\
\textbf{261} & 108 & Risse, Thomas (ed.) & 1995 & Bringing Transnational Relations Back In: Non-State Actors, Domestic Structures and International Institutions (with Thomas Biersteker, Steve Smith) \\
\textbf{25} & 362 & Risse, Thomas (ed.) & 1999 & The power of human rights: International norms and domestic change (with Ropp, Stephen C.; Sikkink, Kathryn) \\
\textbf{69} & 248 & Risse, Thomas & 2000 & ``Let's argue!'': Communicative Action in World Politics. \textit{Int Organ} \textbf{54}, 1 \\
\textbf{258} & 110 & Rodrik, Dani & 1997 & Has Globalization Gone Too Far? \\
232 & 123 & Rogowski, Ronald & 1989 & Commerce and Coalitions. How Trade Affects Domestic Political Alignments \\
\textbf{213} & 132 & Rose, Gideon & 1998 & Neoclassical Realism and Theories of Foreign Policy. \textit{World Polit} \textbf{51}, 152 \\
\textbf{239} & 121 & Ross, Michael L. & 2001 & Does oil hinder democracy? \textit{World Polit} \textbf{53}, 325 \\
\textbf{294} & 72 & Ross, Michael L. & 2004 & What Do We Know about Natural Resources and Civil War? \textit{J Peace Res} \textbf{41}, 337 \\
\textbf{106} & 194 & Ruggie, John Gerard & 1982 & International regimes, transactions, and change -- embedded liberalism in the post-war economic order. \textit{Int Organ} \textbf{36}, 379 \\
\textbf{198} & 139 & Ruggie, John Gerard & 1993 & Territoriality and Beyond: Problematizing Modernity in International Relations. \textit{Int Organ} \textbf{47}, 139 \\
\textbf{161} & 159 & Ruggie, John Gerard & 1998 & Constructing the world polity: essays on international institutionalization \\
\textbf{84} & 223 & Russett, Bruce & 1993 & Grasping the Democratic Peace: Principles for a Post-Cold War World (with William Antholis, Carol R. Ember) \\
\textbf{23} & 384 & Russett, Bruce & 2001 & Triangulating Peace: Democracy, Interdependence, and International Organizations (with John Oneal) \\
\textbf{71} & 246 & Sageman, Marc & 2004 & Understanding Terror Networks \\
\textbf{93} & 214 & Said, Edward W. & 1978 & Orientalism: Western Conceptions of the Orient \\
\textbf{274} & 96 & Sambanis, Nicholas & 2001 & Do Ethnic and Nonethnic Civil Wars Have the Same Causes?: A Theoretical and Empirical Inquiry (Part 1). \textit{J Confl Resolut} \textbf{45}, 259 \\
\textbf{279} & 93 & Sambanis, Nicholas & 2004 & What is civil war? Conceptual and empirical complexities of an operational definition. \textit{J Confl Resolut} \textbf{48}, 814 \\
\textbf{21} & 393 & Schelling, Thomas C. & 1960 & The Strategy of Conflict \\
\textbf{47} & 293 & Schelling, Thomas C. & 1966 & Arms and Influence \\
\textbf{230} & 125 & Schimmelfennig, Frank & 2001 & The community trap: Liberal norms, rhetorical action, and the eastern enlargement of the European Union. \textit{Int Organ} \textbf{55}, 47 \\
\textbf{237} & 121 & Schweller, Randall L. & 1994 & Bandwagoning for Profit: Bringing the Revisionist State Back In. \textit{Int Secur} \textbf{19}, 72 \\
186 & 145 & Scott, James C. & 1998 & Seeing Like a State: How Certain Schemes to Improve the Human Condition Have Failed \\
85 & 223 & Sen, Amartya & 1999 & Development as Freedom \\
\textbf{169} & 155 & Signorino, Curtis S. & 1999 & Tau-b or Not Tau-b: Measuring the Similarity of Foreign Policy Positions. \textit{Int Stud Q} \textbf{43}, 115 (with Jeffrey M. Ritter) \\
\textbf{254} & 113 & Simmons, Beth A. & 2000 & International Law and State Behavior: Commitment and Compliance in International Monetary Affairs. \textit{Am Polit Sci Rev} \textbf{94}, 819 \\
\textbf{195} & 141 & Simmons, Beth A. & 2004 & The Globalization of Liberalization: Policy Diffusion in the International Political Economy. \textit{Am Polit Sci Rev} \textbf{98}, 171 \\
\textbf{280} & 93 & Simmons, Beth A. & 2006 & Introduction: The international diffusion of liberalism. \textit{Int Organ} \textbf{60}, 781 (with Dobbin, Frank; Garrett, Geoffrey) \\
\textbf{117} & 183 & Simmons, Beth A. & 2009 & Mobilizing for Human Rights: International Law in Domestic Politics \\
\textbf{52} & 284 & Singer, J. David & 1972 & Capability Distribution, Uncertainty, and Major Power War, 1820--1965. In: Russett B. (ed.) \emph{Peace, War, and Numbers}, pp.\;19--48 (with Stewart Bremer and John Stuckey) \\
\textbf{217} & 129 & Singer, J. David & 1987 & Reconstructing the correlates of war dataset on material capabilities of states, 1816--1985. \textit{Int Interact} \textbf{14}, 115 \\
227 & 126 & Skocpol, Theda & 1979 & States and Social Revolutions: A comparative analysis of France, Russia, and China \\
\textbf{42} & 316 & Slaughter, Anne-Marie & 2004 & A New World Order: Government Networks and the Disaggregated State \\
\textbf{175} & 152 & Snyder, Jack L. & 2000 & From Voting to Violence: Democratization and Nationalist Conflict \\
\textbf{142} & 167 & Snyder, Jack & 1991 & Myths of Empire: Domestic Politics and International Ambition \\
\textbf{223} & 127 & Spruyt, Hendrik & 1994 & The Sovereign State and its Competitors \\
\textbf{238} & 121 & Stedman, Stephen John & 1997 & Spoiler problems in peace processes. \textit{Int Secur} \textbf{22}, 5 \\
\textbf{78} & 236 & Stiglitz, Joseph & 2002 & Globalization and its discontents \\
\textbf{154} & 163 & Strange, Susan & 1996 & The Retreat of the State: The Difussion of Power in the World Economy \\
299 & 53 & Suchman, Mark C. & 1995 & Managing Legitimacy: Strategic and Institutional Approaches. \textit{Acad Manage Rev} \textbf{20}, 571 \\
83 & 224 & Tarrow, Sidney G. & 1994 & Power in Movement: Social Movements and Contentious Politics \\
\textbf{253} & 114 & Tarrow, Sidney & 2005 & The New Transnational Activism \\
271 & 98 & Tilly, Charles (ed.) & 1975 & The Formation of National States in Europe \\
105 & 197 & Tilly, Charles & 1978 & From Mobilization to Revolution \\
\textbf{73} & 244 & Tilly, Charles & 1990 & Coercion, Capital and European States \\
110 & 190 & Tsebelis, George & 2002 & Veto players: how political institutions work \\
\textbf{273} & 98 & Vachudova, Milada Anna & 2005 & Europe Undivided: Democracy, Leverage, and Integration After Communism \\
\textbf{221} & 128 & Van Evera, Stephen & 1999 & Causes of War: Power and the Roots of Conflict \\
\textbf{143} & 167 & Vasquez, John A. & 1993 & The War Puzzle \\
\textbf{46} & 301 & von Clausewitz, Carl & 1832 & Vom Kriege. Hinterlassenes Werk des Generals Carl von Clausewitz, ed.\;by Marie von Clausewitz (On War) \\
190 & 143 & Wade, Robert & 1990 & Governing the Market: Economic Theory and the Role of Government in East Asian Industrialization \\
\textbf{96} & 209 & Waever, Ole & 1995 & Securitization and Desecuritization. In: On Security. Ed.\;by Ronnie D.\;Lipschutz. Columbia University Press, 1995. pp.\;46--87 \\
\textbf{60} & 263 & Walker, Rob B.J. & 1993 & Inside/outside: international relations as political theory \\
\textbf{29} & 354 & Walt, Stephen M. & 1987 & The origins of alliances \\
\textbf{159} & 160 & Walter, Barbara F. & 2002 & Committing to Peace: The Successful Settlement of Civil Wars \\
\textbf{55} & 276 & Waltz, Kenneth N. & 1959 & Man, the State, and War: A Theoretical Analysis \\
\textbf{1} & 1257 & Waltz, Kenneth N. & 1979 & Theory of International Politics \\
\textbf{184} & 147 & Waltz, Kenneth N. & 2000 & Structural realism after the Cold War. \textit{Int Secur} \textbf{25}, 5 \\
57 & 271 & Weber, Max & 1922 & Wirtschaft und Gesellschaft (Economy and Society) \\
\textbf{111} & 189 & Weinstein, Jeremy M. & 2007 & Inside Rebellion: The Politics of Insurgent Violence \\
\textbf{211} & 132 & Wendt, Alexander E. & 1987 & The agent-structure problem in international-relations theory. \textit{Int Organ} \textbf{41}, 335 \\
\textbf{20} & 394 & Wendt, Alexander & 1992 & Anarchy is what states make of it -- the social construction of power-politics. \textit{Int Organ} \textbf{46}, 391 \\
\textbf{2} & 911 & Wendt, Alexander & 1999 & A Social Theory of International Politics \\
\textbf{91} & 218 & Wheeler, Nicholas J. & 2000 & Saving Strangers: Humanitarian Intervention in International Society \\
\textbf{215} & 130 & Williams, Michael C. & 2003 & Words, images, enemies: Securitization and international politics. \textit{Int Stud Q} \textbf{47}, 511 \\
277 & 93 & Williamson, Oliver E. & 1985 & The Economic Institutions of Capitalism: Firms, Markets, Relational Contracting \\
\textbf{149} & 164 & Wohlforth, William C. & 1999 & The stability of a unipolar world. \textit{Int Secur} \textbf{24}, 5 \\
86 & 223 & Wooldridge, Jeffrey M. & 2002 & Econometric Analysis of Cross Section and Panel Data \\
51 & 288 & Zaller, John R. & 1992 & The Nature and Origins of Mass Opinion \\